\documentclass[useAMS,usenatbib]{mnras}

\usepackage{graphicx}
\usepackage{color}
\usepackage{multirow}
\usepackage{amssymb,amsmath}
\usepackage{cprotect}

\def\aap{A\&A}
\def\apj{ApJ}

\def\apjl{ApJL}
\def\aj{AJ}

\def\pasa{PASA}
\def\mnras{MNRAS}

\def\na{New Astron.}

\def\apjs{ApJS}

\def\ps{$\rm km \,s^{-1}\,kpc^{-1}$}

\def\arcdeg{$^\circ$}
\def\kms{\,km\,s$^{-1}$}

\def\H2{$\rm H_2$}

\def\cc{$\rm cm^{-3}$}
\def\nth{$n_{\rm SF,min}$}

\title[ISM properties in tidal spirals]{Star formation and ISM morphology in tidally induced spiral structures}

\author[A. R. Pettitt et al.]
{Alex R. Pettitt$^{1}$\thanks{E-mail:
alex@astro1.sci.hokudai.ac.jp}, 
Elizabeth J. Tasker$^{1,3}$,
James W. Wadsley$^{2}$,  \and
Ben W. Keller$^{2}$ and
Samantha M. Benincasa$^{2}$ 
\\
$^{1}$Department of Physics, Faculty of Science, Hokkaido University, Sapporo 060-0810, Japan\\
$^{2}$Institute of Space and Astronomical Science, Japan Aerospace Exploration Agency, Yoshinodai 3-1-1, Sagamihara, Kanagawa, Japan\\
$^{3}$Department of Physics and Astronomy, McMaster University, Hamilton, L8S 4M1, Canada\\
}

\begin{document}

\date{\today}

\pagerange{\pageref{firstpage}--\pageref{lastpage}} \pubyear{2017}

\maketitle

\label{firstpage}

\begin{abstract}
{Tidal encounters are believed to be one of the key drivers of galactic spiral structure in the Universe. Such spirals are expected to produce different morphological and kinematic features compared to density wave and dynamic spiral arms. In this work, we present high resolution simulations of a tidal encounter of a small mass companion with a disc galaxy. Included are the effects of gas cooling and heating, star formation and stellar feedback. The structure of the perturbed disc differs greatly from the isolated galaxy, showing clear spiral features that act as sites of new star formation, and displaying interarm spurs. The two arms of the galaxy, the bridge and tail, appear to behave differently; with different star formation histories and structure. Specific attention is focused on offsets between gas and stellar spiral features which can be directly compared to observations. We find that some offsets do exist between different media, with gaseous arms appearing mostly on the convex side of the stellar arms, though the exact locations appear highly time dependent. These results further highlight the differences between tidal spirals and other theories of arm structure.}
\end{abstract}

\begin{keywords}
hydrodynamics - galaxies: spiral - galaxies: structure - galaxies: interactions - ISM: structure 
\end{keywords}

\section{Introduction}

The origin of spiral arms in disc galaxies remains a debated topic despite a great deal of investigation over the past several decades. The density wave picture of spiral arms popularised by the works of \citet{1964ApJ...140..646L}, \citet{1973PASAu...2..174K}, and \citet{1989ApJ...338...78B} still holds as the canonical theory. However, there are many other competing theories that postulate spiral arms behave as dynamical transients \citep{1984ApJ...282...61S,1993A&A...272...37E}, can be created by bar rotation \citep{1980A&A....92...33C,1992MNRAS.259..345A}, tidal interactions \citep{1972ApJ...178..623T,1991A&A...244...52E}, and rotations of non-axisymmetric dark matter haloes \citep{2009ApJ...703.2068D,2012ARep...56...16K}. The emerging paradigm seems to be that there may not be a grand underlying theory for all spirals; and that each theory has its own merits and weaknesses (e.g. \citealt{2010MNRAS.409..396D,2016MNRAS.460.2472B}). It is likely that the theory that applies depends on a number of factors, such as the galaxy's environment, mass model, and evolutionary history. See \citet{2014PASA...31...35D} and \citet{2011MNRAS.410.1637S} for an in-depth and contemporary reviews of spiral structure theories.

Tidal spirals in particular offer a unique test case due to their ability to generate unbarred two-armed spirals with relative ease, as seen in many theoretical works in the literature \citep{1972ApJ...178..623T,1991A&A...252..571D,2000MNRAS.319..377S,2008ApJ...683...94O,2011MNRAS.414.2498S}. As a massive companion passes a host disc galaxy the tidal interaction induces a trailing two-armed spiral in the disc, with a bridge arm formed in the disc side facing the companion, and a tail arm at the opposite side. The progenitors of this interaction could be dwarf galaxies (e.g. \citealt{2010MNRAS.403..625D}) or dark matter subhaloes \citep{2011ApJ...743...35C}, but in general these spirals are predicted to deviate from density wave and dynamic spirals in several ways. Spiral arms in the density wave picture are expected to induce shocks in the gas as it passes through the stellar potential well \citep{1968IAUS...29..453F,1969ApJ...158..123R}. Evidence for which can be seen in the observed thin dust lanes in spiral arms, and high density gas in simulations (e.g. \citealt{2004MNRAS.349..270W,2006MNRAS.367..873D}). Such shocks are believed to be offset from the stellar potential minima. The azimuthal position of these shocks compared to the stellar spiral is dependent on a number of factors such as rotation curve and gas surface density, and result in shocks being both upstream and downstream of the stellar potential \citep{2004MNRAS.349..909G,2006ApJ...647..997S,2008ApJ...675..188W}. However, these shocks are based on the classical density wave picture of spiral arms, and evidence from simulations suggests such shocks in dynamic arms are very different, showing little to no offset between gas and stellar arms \citep{2011ApJ...735....1W,2012MNRAS.426..167G,2015PASJ...67L...4B}. 

The case of tidal arms is somewhat more confusing, as they tend to behave somewhere between dynamic and density wave like structures. Evidence from simulations suggests that they have a radially decaying yet slower than material pattern speed (\citealt{2011MNRAS.414.2498S,2015ApJ...807...73O}). As such, studies in the literature tend to be split on whether they have a negligible \citep{2010MNRAS.403..625D,2011MNRAS.414.2498S} or slight offset between arms of different media (\citealt{2016MNRAS.458.3990P}, hereafter Paper\,1). On the observational side; the most compelling evidence for offset features is in M51, which itself is believed to be a tidally perturbed spiral. Studies looking into numerous tracers in M51 see some offsets in the arms, though the response in each arm seems somewhat different \citep{2013ApJ...779...42S,2017MNRAS.465..460E}. Other galaxies appear to show some offset features, though the picture is hardly consistent \citep{2011MNRAS.414..538K}. Understanding these offsets could be the key to determining the origin of spiral arms in observed galaxies, so determining their nature in a system such as M51 is of great importance. M51 is also seen to have a pattern speed of its spiral arms that decreases with radius in a similar manner as suggested by tidal interaction simulations \citep{2008ApJ...688..224M}.

The main method of studying the time-varying properties of different spiral models is through numerical simulations. Galactic simulations have become increasingly sophisticated in recent years, with ever more complex stellar feedback \citep{2013ApJ...770...25A,2014MNRAS.445..581H}, modern hydrodynamical routines \citep{2014MNRAS.442.1992H,2016MNRAS.460.4382F}, and near pc resolutions \citep{2013MNRAS.436.1836R,2016MNRAS.461.1684F}. Simulations of isolated galaxies with cooling, feedback and star formation are becoming increasingly common in the literature at high resolutions. Studies exist using both simple analytic background potentials \citep{2008PASJ...60..667S,2011MNRAS.417.1318D,2015ApJ...801...33T} and resolved stellar systems \citep{2011ApJ...735....1W, 2012MNRAS.425.2157D,2012MNRAS.421.3488H,2013MNRAS.436.1836R,2015MNRAS.449.2156A,2016ApJ...827...28G}. However, simulations with such complex physics at high resolutions are infrequent in the literature of tidal spirals, and simulations of interactions are often focused on merger events (e.g. \citealt{2009PASJ...61..481S,2013MNRAS.430.1901H}), though early stages of mergers are often progenitors of spiral arms. 
Such mergers tend to cause strong boosts in the star formation activity, up to a 10 times that of the isolated value, though this is a strong function of model configuration (e.g. \citealt{2008A&A...492...31D, 2013MNRAS.430.1901H}). \citet{2016A&A...592A..62G} perform simulations of fly-by and mergers with both grid and particle codes. They found that while different numerical methods tend to give similar global results, the Eulerian models gave a smoother gas disc than the Lagrangian runs, though this could be due to differences in the subgrid physics in each code. \citet{2007A&A...468...61D} performed a huge parameter sweep of simulated interactions and mergers, though with a comparably low resolution. Their interactions show starbursts during interactions, with new star formation mostly due to infall of gas to the galactic centre, and that strong interactions can be effective in subduing star formation activity due to stripping gas from the host galaxy. Recently \citet{2015MNRAS.448.1107M} investigate the spatial distribution of star formation events in an interaction/merger scenario, finding that star formation is boosted only in the initial companion passage, and will then die away to below pre-interaction levels. 

In our previous study on tidal structure (Paper\,1) we performed a parameter sweep of simulations looking into the impact of different companion masses, orbits, and inclinations on the spiral features formed in a disc galaxy. We found evidence of interarm spurs, variations in structure as a function of interaction strength, and longevity of spiral arm pitch angles and pattern speeds. These models were, however, quite simplistic in nature, only including hydrodynamics and gravity and simulating only the warm neutral medium ($T=$10,000K). In this work we investigate the fiducial simulation from Paper\,1, using a much greater resolution and breadth of physics (including cooling, star formation, and stellar feedback) to investigate the nuances of spiral arm features formed in a tidal encounter in a more realistic interstellar medium (ISM) model. Such a study of the locations of star formation, gas morphology and arm structure appear lacking in the literature, with studies instead studying idealised isothermal or $N$-body models, or focusing efforts on modelling mergers and their Gyr scale evolution.

This paper is organised as follows. In Section \ref{sec:numerics} we discuss the numerical models, including the various physical processes, numerical code and simulation initial conditions. In Section \ref{sec:results} we present our results and discussion. Section \ref{sec:isodisc} looks at gas in a static axisymmetric stellar disc potential, which acts as a test-case for the ISM models. Section \ref{sec:livedisc} presents calculations using a live $N$-body disc before the introduction of the companion, and Section \ref{sec:pertdisc} the perturbed live galactic disc. We conclude in Section \ref{conclusions}.

\section{Numerical Simulations}
\label{sec:numerics}
\subsection{Hydrodynamics and Gravity}
Simulations were performed using the $N$-body+smoothed particle hydrodynamics (SPH) code \textsc{gasoline} \citep{2004NewA....9..137W}, a descendant of the gravity tree-code \textsc{pkdgrav} \citep{2001PhDT........21S}. Gravity is solved using a binary tree, and the system integrated using a kick-drift-kick leapfrog. Self gravity is active for all components, using a fixed gravitational softening of 10pc. We choose to use the standard M4 cubic spline kernel with 64 neighbours.

\subsection{Cooling}
The gas in these simulations is initially isothermal with a temperature of 10,000K, and is allowed to cool using a tabulated cooling function from \citet{2010MNRAS.407.1581S}, which spans a temperature and density ranges of $1{\rm K}\le T \le 1\times10^9 {\rm K}$ and $1\times10^{-9}{\rm cm ^{-3}}\le n \le 1\times10^4 {\rm cm^{-3}}$. Similarly to \citet{2016MNRAS.462.3053B}, we use a metal depletion factor of 0.6 to ensure the creation of a two-phase ISM. In addition to metal cooling, we also include UV and photoelectric heating. However, unlike in \citet{2016MNRAS.462.3053B} and \citet{2009ApJ...700..358T} who use a radial dependence tailored to the Milky Way, we adopt a constant factor approximately equal to that at the solar radius. This is due to the galaxies being modelled not necessarily as Milky Way analogues, with the tidally perturbed case having disc scalelengths and surface densities that are likely to vary strongly with time.

\begin{figure}
\centering
\resizebox{1.\hsize}{!}{\includegraphics[trim = 10mm 0mm 0mm 0mm]{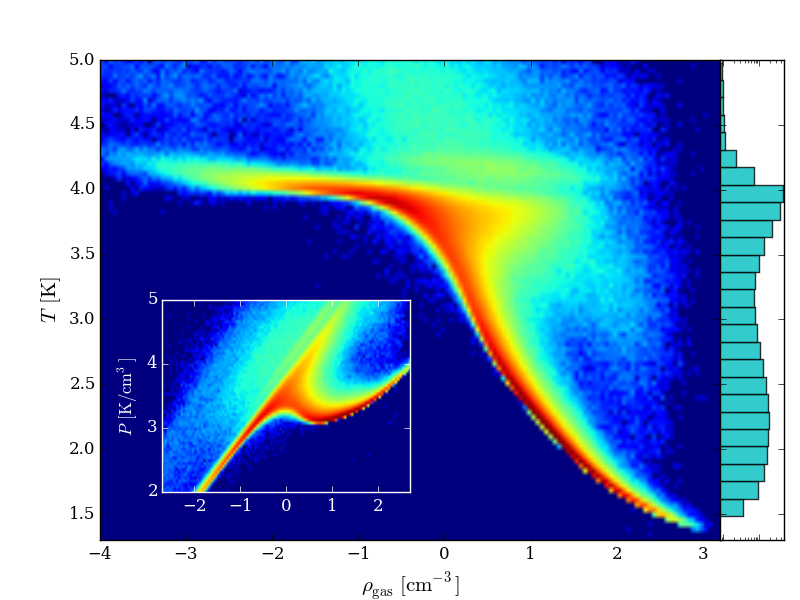}}
\caption{Temperature-pressure-density diagrams for the cooling/heating functions used in this study. The main frame shows a density map of the density-temperature profile of all gas particles, with a histogram of temperature to the right hand side. The small insert shows the pressure-density profile, clearly showing the two-phase nature of the ISM. The data here are from the static axisymmetric disc calculation after 500Myr of evolution (see Sec. \ref{ICiso}).}
\label{phasecurve}
\end{figure}

These values were chosen to give a thermal profile that supposedly mimicked that of the ISM, namely of a two-phase equilibrium \citep{2003ApJ...587..278W}. An example of our phase curve is shown in Figure\;\ref{phasecurve}. This figure clearly shows the two phase medium with populations peaking at around 10000K and 200K for the warm and cold neutral media, and reaching temperatures as low as 20K. While the cooling table goes to lower temperatures, increased resolution and molecular chemistry are needed to model this correctly, thus we enforce a 10K lower limit in these calculations. A pressure floor prevents artificial fragmentation below resolution scale, and is calculated as $P_{\rm floor}=3\rho^2 l^2G $, where $l={\rm max}(h,\varepsilon_{\rm grav})$ is the spatial resolution \citep{2009MNRAS.397L..64A} set to the maximum of smoothing lengths ($h$) and gravitational softening ($\varepsilon_{\rm grav}$).

\subsection{Star formation}
We use standard subgrid single-stellar-population star formation models \citep{1992ApJ...391..502K,1996ApJS..105...19K,2006MNRAS.373.1074S}. The main parameters governing star formation are $c_*$, the star formation efficiency, and \nth{}, the density threshold. We use a value of \nth{}=100\cc{}, $c_*=6\%$, and a maximum temperature for star formation of $T_{\rm SF, max}=300$K in all the simulations presented here. Several studies have investigated the impact of changing these parameters, so we simply adopt similar values to other studies at this scale \citep{2006ApJ...641..878T,2008PASJ...60..667S,2008ApJ...673..810T}. For each star formation event only a single star is created from the parent gas particle, meaning a 100\% conversion of a gas particle to a single star particle. This ratio is often chosen to be around a third of the gas particle mass \citep{2005MNRAS.363.1299O,2007MNRAS.374.1479G,2016MNRAS.455..920W} giving a good balance between dynamical complexity and resolution. However, here we choose a 1:1 ratio to ensure particle masses are approximately equal, regardless of type, to lessen errors in the SPH method when using unequal particle masses.

\subsection{Feedback}
Stellar feedback is included from SNII and SNIa supernova, and winds, in a blast wave feedback approach with a standard cooling shut-off for each gas particle in the blast wave (see \citealt{2006MNRAS.373.1074S} for details). We adopt a feedback energy of Type II SN, $E_{SN}$
 of $10^{50}\,{\rm erg}$. This gives reasonable star formation rates (SFR) when combined with our value of $c_*$, which gives a SFR of the order of $\rm 1-10\;M_\odot\;yr^{-1}$

We also experimented with two other feedback schemes, the constant cooling delay feedback of \citet{2013ApJ...770...25A} and the superbubble feedback of \citet{2014MNRAS.442.3013K}. However, while each proposes improvements over the blastwave model, we found strong ring features in the constant cooling delay runs (see also \citealt{2016MNRAS.462.3053B}) and muted morphological features in the superbubble runs (see also \citealt{2015MNRAS.453.3499K,2016ApJ...830L..13M}). As such we employed the widely adopted blast wave model due to the more responsive morphological features and star formation rates.

\subsection{Simulation parameters}
We describe three different simulation models. The Static model where gas is embedded in a static axisymmetric potential, the Live model where an $N$-body system of particles represents the old stars and halo, and the Pert model where the Live model experiences the tidal interaction of a passing companion.

\subsubsection{Static stellar disc}
\label{ICiso}

While the main part of this investigation focuses on the tidally perturbed system, we also include a much simpler model to act as a control group. The purpose of this run is to establish a benchmark against the live and perturbed systems (see Sec. \ref{IClive} and \ref{ICpert}), where the live stellar systems will act to churn and disturb the ISM. This simple model consists of a gas disc embedded in a simple logarithmic potential to represent a Milky Way-like flat rotation curve \citep{1987gady.book.....B} of the form:
\begin{equation}
\Phi(r,z) = \frac{1}{2} V_0^2 \log\left[   r^2 + R_c^2 +z^2/q_z^2    \right]
\end{equation}
where the radial core radius is $R_c=1$kpc, the vertical scaling factor is $q_z=0.7$, and flat rotation speed is $V_0=220$\kms{}.
Gas is set radially in a exponential disc with scalelength $r_0=7$kpc (approximately double the Milky Way's stellar disc), and vertically following a ${\rm sech}^2(z/z_0)$ profile where $z_0=0.4$kpc. The system has a resolution of 4\,million gas particles of combined gas mass of $4\times10^9 M_\odot$ out to a radius of 10kpc. Gas is initialised at 10,000K with a solar metallicity (0.013).
This setup is similar to previous works of gas embedded in a galactic potential, such as \citet{2008PASJ...60..667S}, \citet{2011MNRAS.417.1318D}, \citet{2015ApJ...801...33T} and \citet{2016MNRAS.462.3053B}. The system is then allowed to evolve with all physics active from initialisation.

\subsubsection{Isolated multicomponent system}
\label{IClive}
Here we use the same mass model as Paper\,1 where the gas is set using the \textsc{magalie} initial conditions generator \citep{2001NewA....6...27B}.
We utilise a much higher resolution version of the model in Paper\,1, with a resolution of $3\times10^6$ gas particles, $3\times10^6$ old disc stars, $1\times10^5$ bulge stars, and $1\times10^6$ halo particles. The rotation curve and initialisation are identical to Paper\,1. This disc to halo mass ratio was chosen so that the system has a high arm number in isolation, and impedes bar formation. The gas disc setup is slightly different from the previous as we utilise an extended gas disc out to 20kpc; with double the radial scalelength of the gas disc, since gas discs are observed to be flatter and extend further than their stellar counterparts \citep{1986ApJ...308..600T,1998gaas.book.....B}. We use a gas disc of mass of $6\times10^9 M_\odot$, which gives a similar surface density to the Static model (Sec. \ref{ICiso}), that is, of the order of $10M_\odot {\rm pc^{-2}}$.

We allow this live system to relax isothermally for 600\,Myr (approximately two rotations) before activating the cooling, star formation, and feedback. This allows for the gas disc to relax, and the system to begin forming spiral like features. 

\subsubsection{Perturbed multicomponent system}
\label{ICpert}
The perturbing computation is the same setup as Paper\,1 using a resolved companion composed of 10,000 point mass particles, giving a mass resolution similar to that of the dark matter halo of the host galaxy. The total mass of the companion is $2\times 10^{10}M_\circ$ and is initialised as a Plummer sphere using \textsc{magalie} \citep{2001NewA....6...27B}. The companion contains no gas component, and is made of an inert old stellar population (no feedback), but could equivalently be assumed to be a dark matter subhalo. The companion is introduced to the system 200\,Myr after the star formation, feedback, and cooling are activated in the multicomponent disc (i.e. 800\,Myr after initialisation), so that the star formation has reached a pseudo-equilibrium state before the companion is introduced. The companion approaches a near-parabolic orbit in the $x-y$ plane, with an initial position of $y=40$kpc and a velocity magnitude of 190\kms. This gives a closest approach of approximately 20kpc from the centre-of-mass of the host galaxy, and a velocity difference with a magnitude of 275\kms{} (giving an orbital frequency of 14\ps{}). The mass ratio of the companion to the host galaxy is 0.092, or equivalently 2/3 of the host stellar disc. These values correspond to a strength parameter for the interaction of $S=0.1$ \citep{1991A&A...244...52E}. See Paper\,1 for further details regarding the perturbed morphology in a simplified hydrodynamics+gravity only case.

\section{Results and Discussion}
\label{sec:results}

\subsection{Gas disc in an axisymmetric static potential}
\label{sec:isodisc}
\begin{figure}
\centering
\resizebox{1.\hsize}{!}{\includegraphics[trim = 0mm 0mm 20mm 10mm]{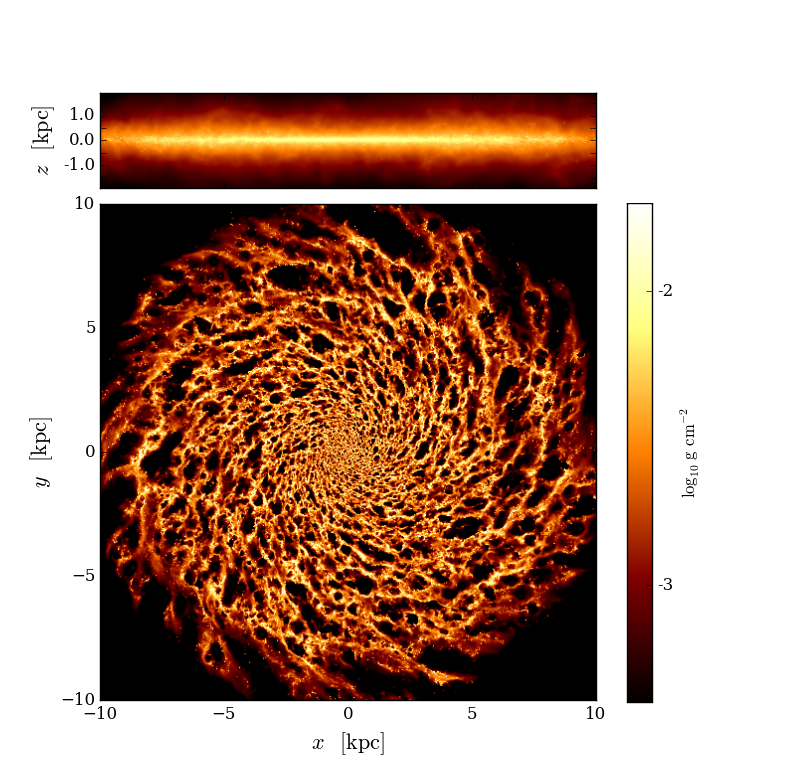}}
\resizebox{0.95\hsize}{!}{\includegraphics[trim = 20mm 0mm 15mm 5mm]{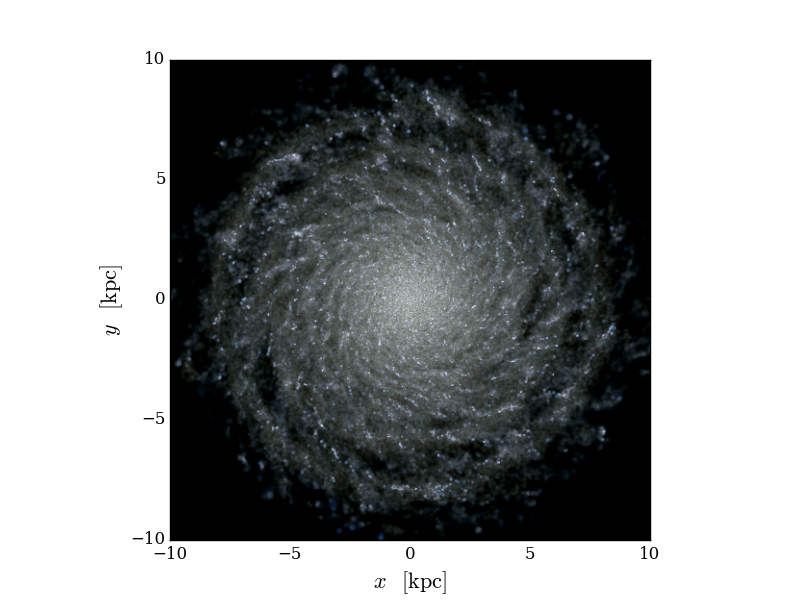}}
\caption{Gas (top) and stellar (bottom) material in Static simulation which includes a static, axisymmetric background potential. The gas is shown as an integrated column density map through the entire domain, while the stellar map shows mock three-colour image of the stars (rbg) dictated by the ages of the stars formed in the simulation. Both images were made using the \textsc{pynbody} package \citep{2013ascl.soft05002P}.}
\label{StatRenders}
\end{figure}

The simplest model here is Static model, shown in Figure\;\ref{StatRenders}. The top panels show a top-down gas density render and a vertical profile, and the bottom panel shows a mock stellar image, both after 400Myr of evolution. The mock stellar image was made using the \textsc{pynbody} package \citep{2013ascl.soft05002P} using the metallicity and ages of newly formed stars to calculate luminosities in \emph{i-v-u} bands mapped to \emph{r-g-b} colour channels. The disc shows a flocculent spiral in the gas. A power mode analysis of spiral features showed that no spiral mode is dominant (up to $m=6$) in the stars or the gas. Several cavities are formed by the stellar feedback, slightly increasing in size with increasing disc radius. The stellar image shows weaker axisymmetric structure, with new (bluer) stars showing the clearest asymmetry, and brightness dropping off to the disc edge (as expected form the exponential gas disc). 

\begin{figure}
\centering
\resizebox{1.\hsize}{!}{\includegraphics[trim = 0mm 0mm 0mm 10mm]{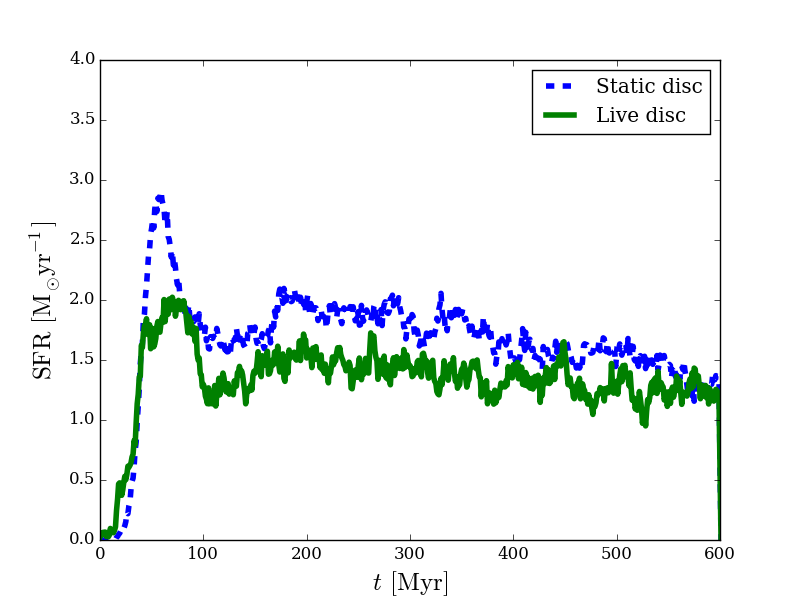}}
\caption{Star formation histories for the Static (blue dashed) and Live )solid green) stellar disc simulations. Rates are calculated across the entire galactic disc.}
\label{SFH}
\end{figure}

The star formation history (SFH) for this and the Live calculation (see Section\;\ref{sec:livedisc}) are shown in Figure \ref{SFH}. The SFH shows a clear peak as the gas initially cools and collapses from the initial 10,000K, which causes this burst in star formation. This then levels off to a normal rate of about 1.5$M_\odot{\rm yr^{-1}}$ for the duration of the simulation. This is similar to what is seen in other disc simulations \citep{2008PASJ...60..667S,2011MNRAS.417.1318D,2013MNRAS.432.2647H,2016MNRAS.462.3053B,2017MNRAS.466.1093G,2017MNRAS.466...11R} and is a result of the initial fragmentation and cooling of the disc. A mild transient ring of old stars seen is visible in the mock stellar image (at a radius of about 8kpc) that has propagated from the centre. This ring was much clearer in simulations with a greater feedback efficiency, and is likely a result of the initial fragmentation from the idealised isothermal initial conditions.

\begin{figure}
\centering
\resizebox{1.\hsize}{!}{\includegraphics[trim = 0mm 0mm 0mm 10mm]{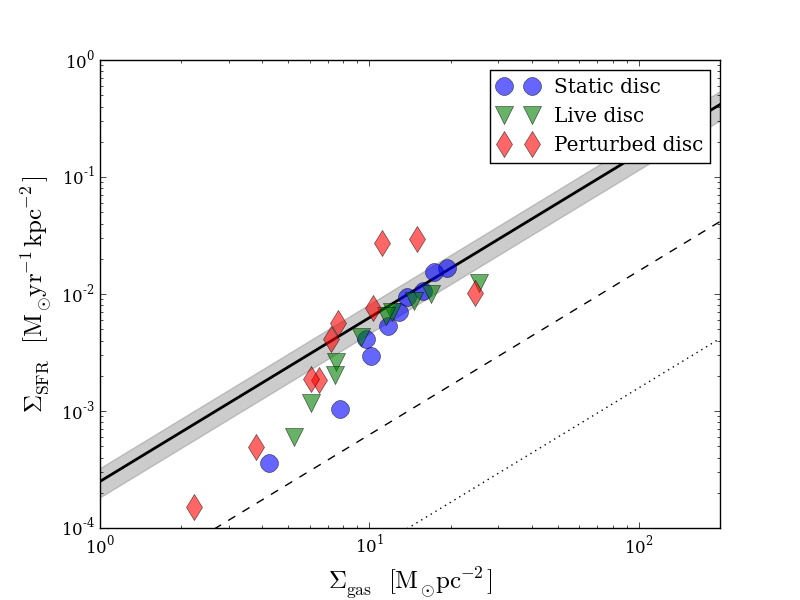}}
\caption{The KS relation for star formation in disc galaxies for the calculations shown here after 600Myr of evolution with star formation, cooling and feedback active. The shaded grey line shows the observationally motivated KS law with error bounds denoted by the shaded region, with laws of 10\% and 1\% the fiducial formation efficiencies denoted by the dashed and dotted lines respectively. Each point denotes a radial annulus bin of width 1kpc.}
\label{KS1}
\end{figure}

A measurable characteristic of well observed disc galaxies is that between gas surface density, $\Sigma_{\rm gas}$, and star formation surface density, $\Sigma_{\rm SFR}$. This Kennicutt-Schmidt relation (KS-relation; \citealt{1959ApJ...129..243S,1998ApJ...498..541K}) takes the basic form of;
\begin{multline}
\Sigma_{\rm SFR} = (2.5\pm 0.7) \times 10^{-4} M_\odot\;{\rm yr^{-1} }\; {\rm kpc^{-2}} \\ \times \left(  \frac{\Sigma_{\rm gas}}{1M_\odot\; {\rm pc^{-2}}}  \right)^{1.4\pm0.15}
\label{KSeq}
\end{multline}
on a galactic scale, whereas smaller scale star formation follows somewhat different relations \citep{2008AJ....136.2846B}. We plot this relation for the Static simulation in Figure \ref{KS1} as blue circles. We plot equation \ref{KSeq} as a black line with error bounds represented as shaded region, and 10\% and 1\% of this rate as dashed and dotted lines respectively. The points for the simulation are calculated at 1kpc annuli over the entire disc, with the highest surface density point being the disc centre. The measured points follow the KS-relation reasonably well, and tail off at low surface densities similar to the sub-kpc measurements of \citet{2008AJ....136.2846B}.

The purpose of this model is to act as a control test for the cooling (Fig. \ref{phasecurve} uses the data from this model), feedback and star formation on these scales, and many of the multitude of parameters involved were chosen to give sensible phase curves, SFR and morphology.

\subsection{Isolated live galactic disc}
\label{sec:livedisc}

\begin{figure}
\centering
\resizebox{1.\hsize}{!}{\includegraphics[trim = 0mm 0mm 20mm 10mm]{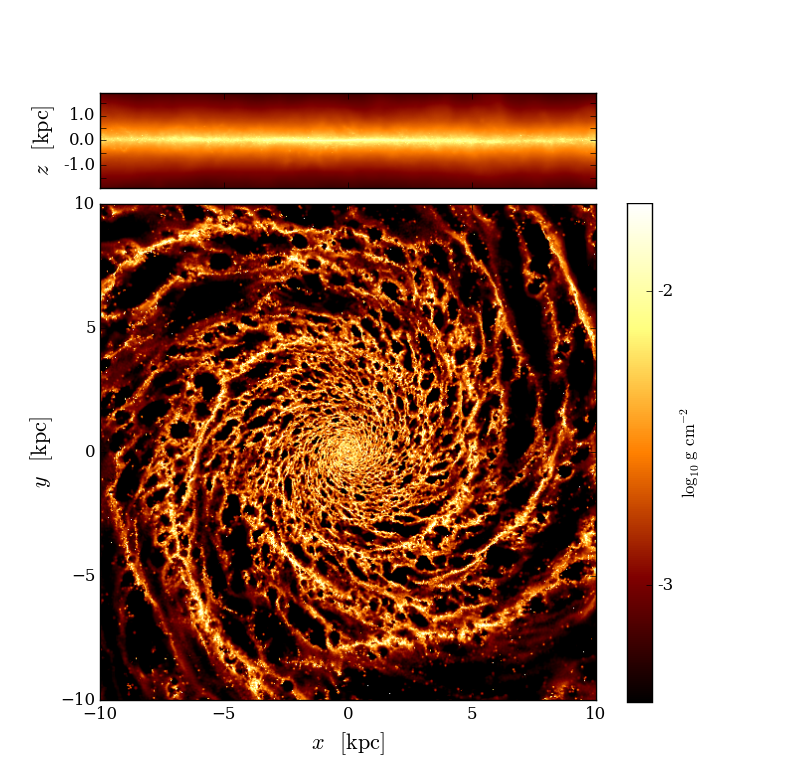}}
\resizebox{0.95\hsize}{!}{\includegraphics[trim = 20mm 0mm 15mm 5mm]{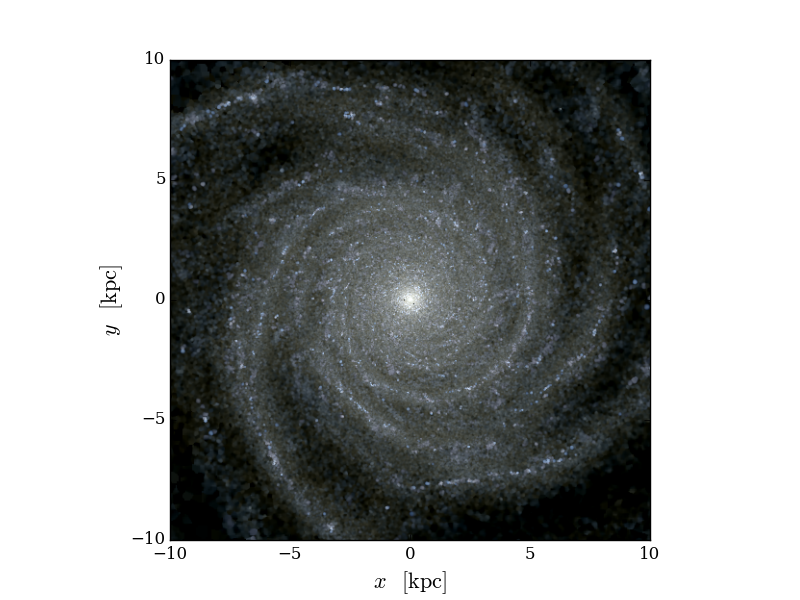}}
\caption{As Figure\;\ref{StatRenders} but for the live isolated disc. Note that the stellar mock image includes the old stellar population.}
\label{LiveRenders}
\end{figure}

In Figure \ref{LiveRenders} we show a similar plot to Figure \ref{StatRenders}, but for the Live calculation after 400Myr of the activation of cooling, star formation and feedback. In this case the mock stellar image is made of both old and new stellar components, where the old stars (those present at $t=0$) have been assigned an age corresponding to the simulation duration and solar metallicity (the old population is responsible for the central bulge emission). The disc has developed a clear multi-armed spiral component in both the gas and the stars. At the time stamp-shown the dominant mode in the mid disc is $m=3$, with the dominant arm mode increasing with increasing radius in accordance with swing amplification theory (see other recent simulations: \citep{2015MNRAS.449.3911P,2015ApJ...808L...8D}). Once again the gas shows clear cavities caused by clustered stellar feedback, showing a marked difference to the stellar density. High density gas arms tend to be coincident with the stellar arms, this will be discussed further in Section \ref{Sec:offset}.

Figures \ref{SFH} and \ref{KS1} include the data from this model, shown as the green line and triangle points respectively. The SFH shows a similar initial burst as the Static calculation, though relatively shallower compared to the equilibrium rate. This calculation has a steadier SFH at late times, maintaining a rate of about $1.3\rm M_\odot{\rm yr^{-1}}$, whereas the Static model shows a gradual decay on the Gyr time-scale. The points in Figure \ref{KS1} show only minimal difference to the Static model, though this is on a logarithmic scale compared to the data in Figure \ref{SFH}. The slightly higher values in the outer disc result from the spiral presence triggering star formation, whereas the Static disc has only gas-self gravity and a much lower mass young stellar disc to seed collapse. The lower value for the central radii is due to the random bulge star orbits stabilising the disc to collapse, compared to the Static disc where only rotational shear stands in the way of star formation.

\subsection{Perturbed live disc}
\label{sec:pertdisc}

\subsubsection{Basic morphology}

\begin{figure}
\centering
\resizebox{1.\hsize}{!}{\includegraphics[trim = 0mm 0mm 20mm 10mm]{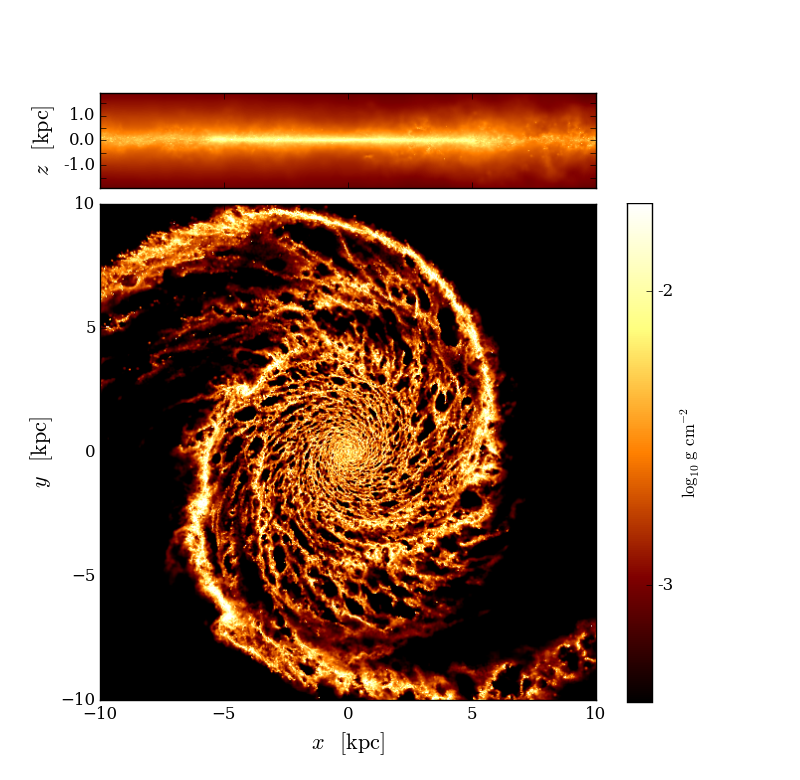}}
\resizebox{0.95\hsize}{!}{\includegraphics[trim = 20mm 0mm 15mm 5mm]{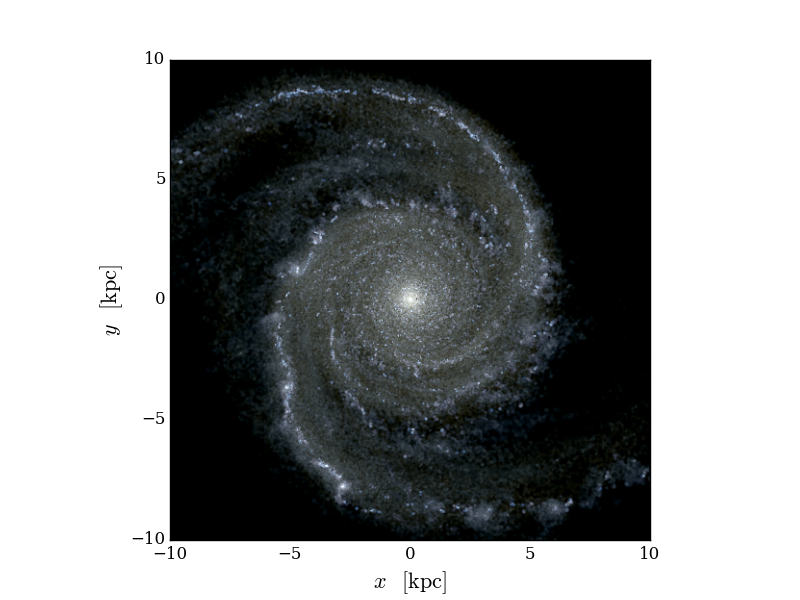}}
\caption{As Figure\;\ref{LiveRenders} but for the perturbed galactic disc, 400Myr after the inclusion of the companion (approximately 200Myr after the closest approach). }
\label{PertRenders}
\end{figure}

Figure \ref{PertRenders} shows the stars and gas in the perturbed galactic disc 450Myr after the introduction of the companion into the system. The central disc is very similar in appearance to the isolated live disc (Fig. \ref{LiveRenders}), and the tidally induced two-armed component dominates outside of approximately 3kpc. The lack of inner spiral features is the product of the mass model and rotation curve. A massive bulge was chosen to inhibit bar formation, which allows for the study of the tidal spiral without contamination by bar-driven spiral arms. As such the high $Q$-barrier that suppresses bar formation also inhibits the penetration of the spiral arms to the inner disc. We will be studying the impact of different mass models on the tidal response in a future study (Pettitt et al., in preparation). The mock stellar image shows regions of fresh star formation predominately along the spiral arms, coincident with the high density gas. 

\begin{figure*}
\centering
\resizebox{1.\hsize}{!}{\includegraphics[trim = 20mm 0mm 0mm 0mm]{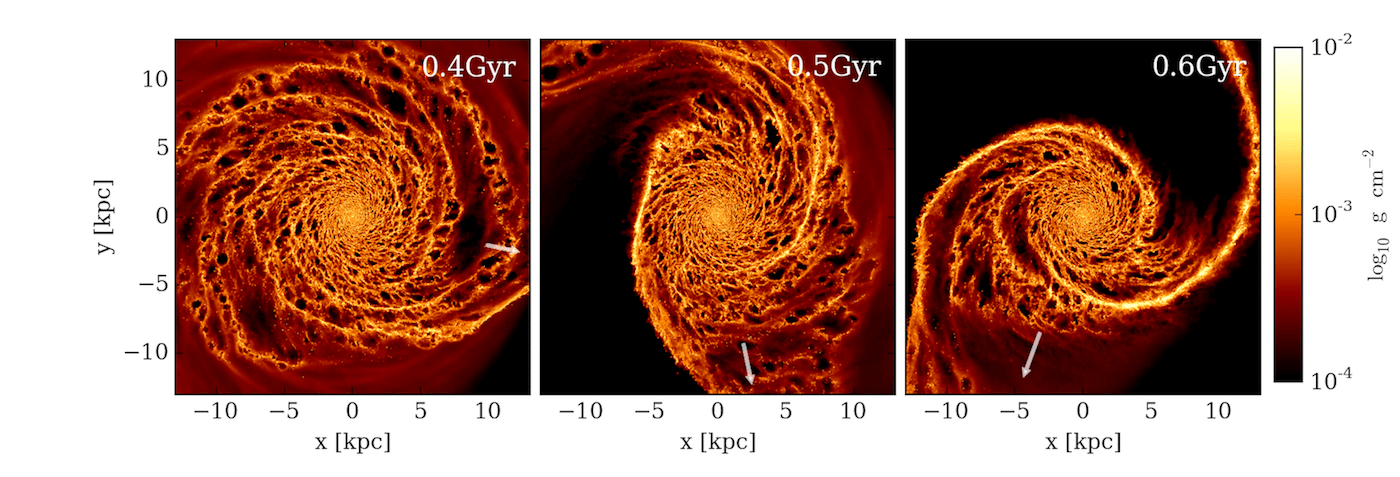}}
\resizebox{1.\hsize}{!}{\includegraphics[trim = 20mm 0mm 0mm 10mm]{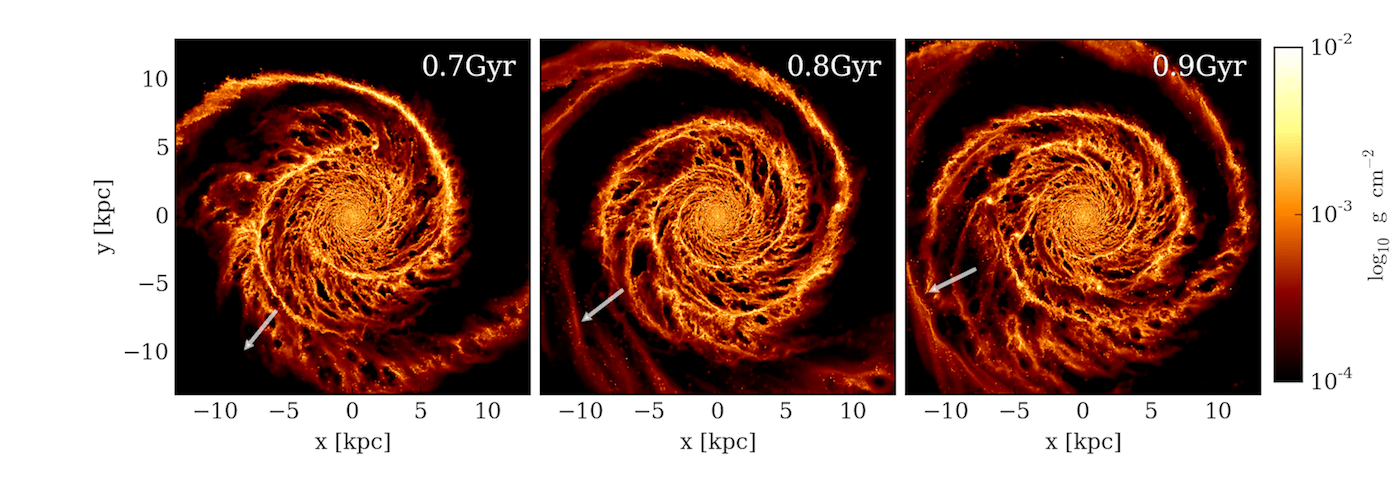}}
\resizebox{1.\hsize}{!}{\includegraphics[trim = 20mm 0mm 0mm 10mm]{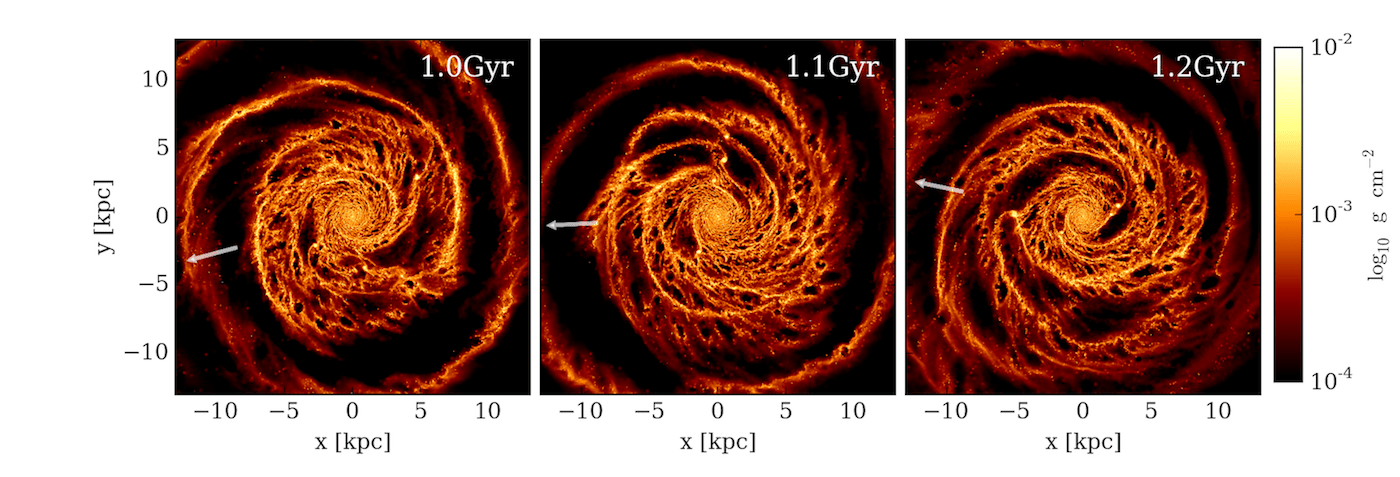}}
\caption{Time series top-down density renders of gas in the perturbed galactic disc for the immediate 800Myr after the closest approach of the companion. The white arrow points in the direction of the companion, where the length is proportional to the distance away from the main galaxy.}
\label{PertEvol}
\end{figure*}

A time-series of the perturbed gas disc is shown in Figure \ref{PertEvol} from the time of closest approach (20kpc from galactic centre) to 800Myr after. The white arrows point to the position of the companion, with a length proportional to the distance from the host centre of mass. Directly after the closest approach a strong bridge is made between the companion and the host, with the tail formed on the opposite side of the disc. As the disc winds up the gas experiences a widespread collapse along the spiral arms, which in a few extreme cases causes the creation of large dense knots (e.g. 0.7 and 1.1Gyr). Once formed, these knots orbit around the disc at the material pattern speed and slowly migrate towards the galactic centre over the course of many rotations. The spiral arms themselves only appear to last for 2-3 disc rotations. After this, they stray from a regular logarithmic-spiral structure, wounding up, as well as disrupted by feedback and clump passage though the arms.

\subsubsection{Star formation history}

\begin{figure}
\centering
\resizebox{1.\hsize}{!}{\includegraphics[trim = 0mm 0mm 0mm 10mm]{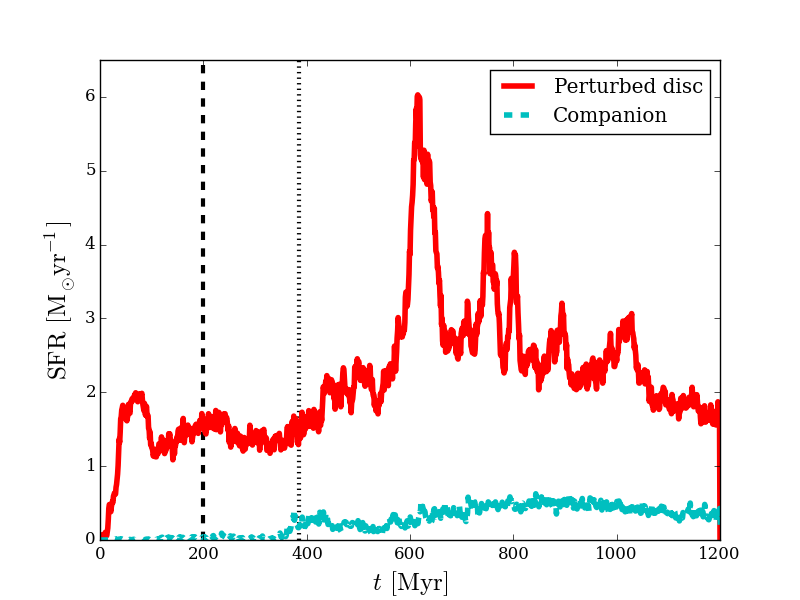}}
\caption{Star formation histories for the perturbed live stellar disc (red solid line) and the companion (cyan dashed line). The vertical dashed line shows the time when the companion is introduced to the system, and the dotted line indicates the time of the closest approach.}
\label{SFHcomp}
\end{figure}

Figure \ref{SFHcomp} shows the SFH of the perturbed galactic disc in red over a period of 1.2Gyr. The vertical dashed line shows the time when the companion is introduced to the simulation, and prior to this the SFH has the same form as the unperturbed live disc in Figure \ref{SFH}. The dotted vertical line indicates the closest approach of the companion. Note the peak in star formation at 50Myr from the initial fragmentation is identical to that in Fig. \ref{SFH}. Closest approach marks the beginning of a steep rise in star formation activity, with the rate reaching a peak of $6M_\odot {\rm yr^{-1}}$, approximately four times that of the unperturbed rate. This peak occurs at 650Myr, shortly after the 0.6Gyr density render in Figure \ref{PertEvol}. The top-down map suggests that the tail arm is reaching higher densities compared to the bridge arm and interarm disc. This is caused by pile-up/crowding of orbits in the tidal tail, while bridge orbits are disrupted by continued presence of the companion. This will be discussed in detail later in this section, and has been reported in previous works (e.g. \citealt{2000AJ....120..630E,2008ApJ...683...94O}). The near simultaneous collapse of the gas in the tail arm is the cause of the peak in at 650Myr in Figure \ref{SFHcomp}, and two high density clumps resulting from this collapse can be seen in the interarm regions in the 0.7Gyr time-stamp in Figure \ref{PertEvol}. The next burst in star formation is seen around 750Myr, which appears to coincide with the collapse of the bridge arm, shortly after the 0.7Gyr render in Figure \ref{PertEvol}. This suggests that the arms have multiple times of collapse, with the tail collapsing before the bridge, and then experiencing another burst event at a later time. The undulating trend then continues throughout the simulation, but progressively weaker with time until the SFR becomes comparable with the unperturbed state. 

\begin{figure}
\centering
\resizebox{.8\hsize}{!}{\includegraphics[trim = 10mm 0mm 20mm 10mm]{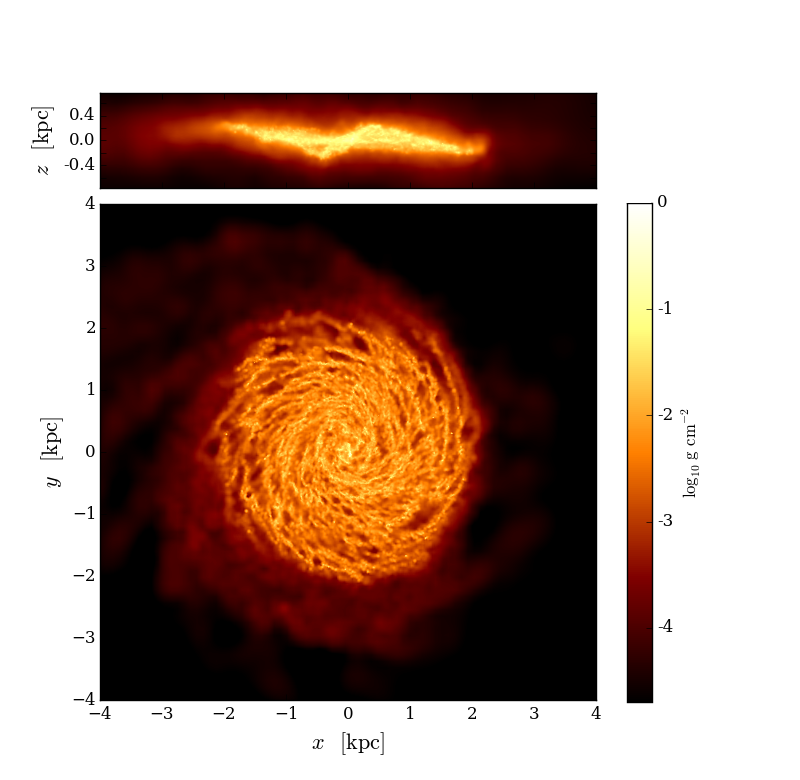}}
\caption{Surface density render and vertical profile of the gas in the companion, 800Myr after the closest approach with the host galaxy. A flocculent disc has been accreted on to initially gas-free structure.}
\label{CompRend}
\end{figure}

The companion stripped away gas from the primary during its passage. Figure\;\ref{CompRend} shows the companion 800Myr after the closest approach. A small disc of gas has been accreted on to the quiescent and axisymmetric stellar companion, which then proceeds to form its own new star particles as it moves away from the host system. 
At the time of Figure\;\ref{CompRend} the companion has captured approximately $3.5\times 10^8 {\rm M_\odot}$ of gas from the host galaxy, though contains $2.5\times 10^8 {\rm M_\odot}$ of new stars that have formed at least partially from captured gas (as well as $7\times 10^8 {\rm M_\odot}$ of captured old stars from the host stellar disc). The cyan dashed line in Figure \ref{SFHcomp} shows the SFH inside the companion. It shows a clear increase after the closest approach and then settles into a semi-steady state for nearly 600Myr. The minor increase before closest approach is due to companion capturing stars that have formed on the outskirts of the host disc and are located within the companion post-interaction. Note however that mass transfer to companions is a complex process, and will vary strongly on the impact parameter, galaxy to companion mass ratio, and the orbital inclination in both fly-bys \citep{1992ApJ...399...29W} and collisions \citep{1997ApJS..113..269S}.

\begin{figure}
\centering
\resizebox{0.9\hsize}{!}{\includegraphics[trim = 8mm 0mm 10mm 0mm]{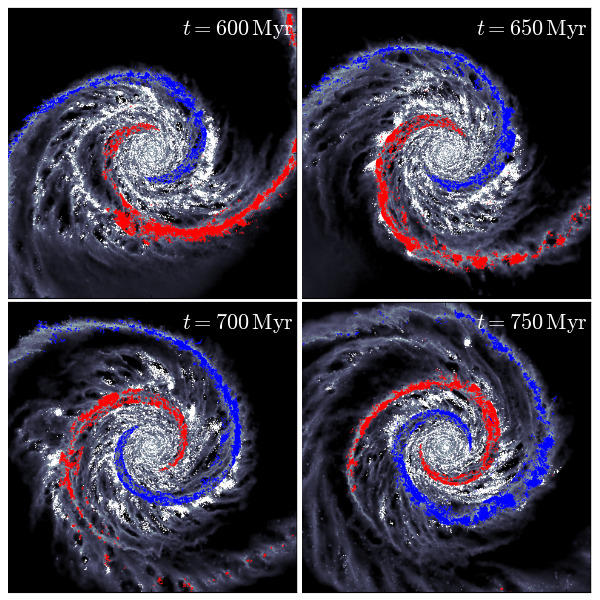}}
\caption{The sites of new star formation in the perturbed disc at four times shortly after companion closest approach. Newly formed stars are coloured red and blue for the tail and bridge arms respectively, and white for interarm stars in the same age bandpass as those shown for the arms. A density render of the gas is plotted under the stars for reference.}
\label{SFRarmstop}
\end{figure}

We select arm regions from the main disc to further investigate the star formation within each individual arm. In Figure \ref{SFRarmstop} we show the perturbed disc at four time-stamps where we have highlighted the newly formed stars, defined here as having ages $<$50Myr, as the coloured points overplotted on a density render of the gas. Blue and red points indicate stars formed within the bridge and tail arms respectively, and white indicates the interarm young stars. Arms are defined by fitting a two-armed log-spiral to the old stellar disc (see Sec. \ref{Sec:offset} for details), with the arm region encompassing $4\pi/10$ radians. There is still ample star formation in the interarm regions, though the star formation in the arms is heavily clustered and contains approximately 50-90\% of the star formation activity, depending on the epoch.

\begin{figure}
\centering
\resizebox{0.95\hsize}{!}{\includegraphics[trim = 10mm 5mm 0mm 5mm]{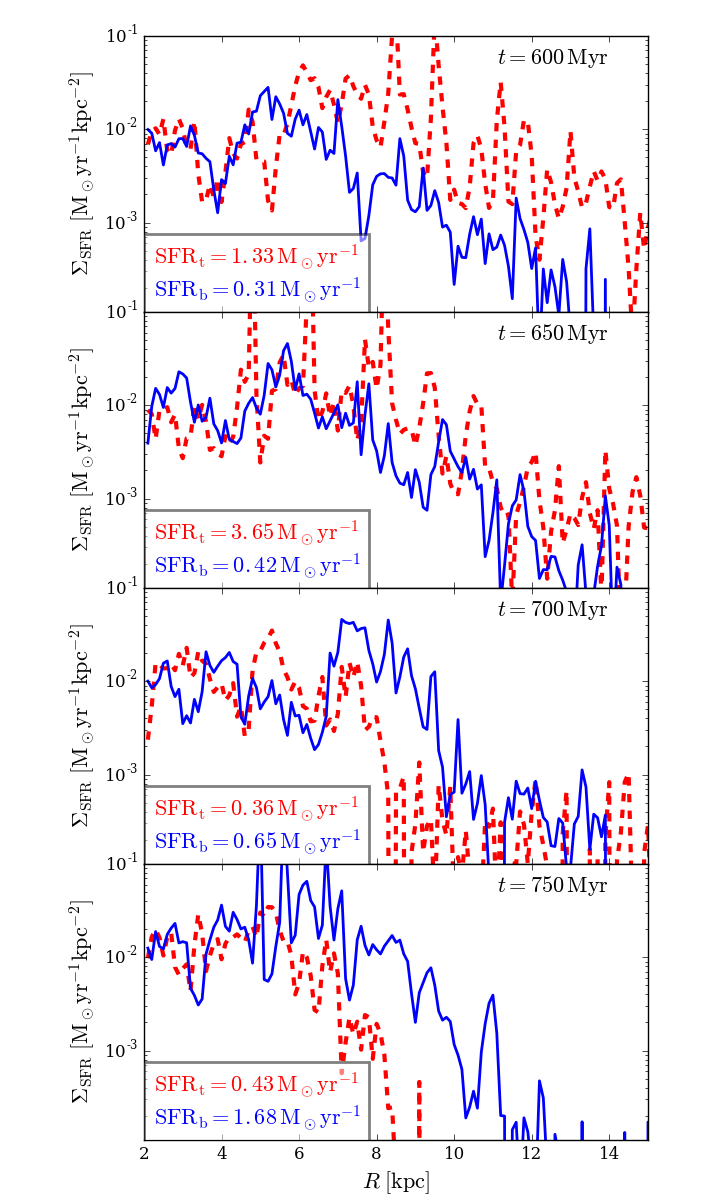}}
\caption{The surface SFR in the arms in each of the time-stamps in Figure\,\ref{SFRarmstop}. Red and blue lines represent the tail and bridge arms, respectively. The total SFR in each arm is shown in the bottom left of each panel.}
\label{SFRarms}
\end{figure}

The star formation surface density for each of the arm regions is plotted as in Figure \ref{SFRarms} at the same four different times as Figure \ref{SFRarmstop}, with matching colour codes for each arm. The total SFR for each arm is given in the bottom left insert. The $\Sigma_{\rm SFR}$ drops towards the disc edge, following the gas surface density profile. The trends between the two arms follows what was postulated by inspection of the undulating nature of the global SFR (Fig. \ref{SFHcomp}). The arms appear to swap in terms of star formation efficiency, with the tail arm experiencing the stronger star formation activity at 600Myr, and the bridge arm showing stronger rates at 750Myr. The star formation activity also appears to retreat towards the galactic centre over time, owing to the spiral features gradual winding up. This can be seen from the top-down maps of Figure \ref{SFRarmstop}, where we can see the radial extent of new stars being smaller in the 650Myr map compared to that of 600Myr (both for arm and interarm regions).

The cause of this offset in collapse times and SFH between the two arms is due to two effects. The first is the nature of the orbits in each arm. In \citet{2008ApJ...683...94O} the authors showed that the tail arm is made of collective overlapping of particle orbits on the far side of the disc (see their Fig. 5). Whereas the orbits in the bridge have been elongated in a teardrop-like feature and are experiencing forced oscillations due to the close proximity of the companion. The infalling gas then flows through the density-wave like bridge feature in a steady stream that extends to the perturber. Also note the much sharper arm-interarm contrast in the tail arm where the orbits of almost all particles have become coincident, compared to the bridge arm which still has a large amount of upstream gas ready to fall in the bridge region and a weaker arm-interarm contrast. 
 \citet{2000AJ....120..630E} also saw such tidal pile-ups in the tail arms of simulations, and inferred that such mechanisms were occurring in the interacting galaxy IC\,2163 (see their Fig. 2). Recent observations of this galaxy made using the Atacama Large Millimeter Array also show a strong pile-up of molecular gas into `eyelid'-shaped arm features that also display asymmetries between bridge and tail regions \citep{2016ApJ...831..161K}.

The second reason for this SFH offset is that much of the bridge arm gas mass has transferred to the perturber. In \citet{2007A&A...468...61D} strong interactions tended to disrupt and mute star formation on a galactic scale due to stripping of the gas disc by the companion. In our models the tidal forces are not strong enough to cause widespread stripping, but have instead transferred a significant amount of the gas budget via the bridge tidal arm (seen in Figure \ref{CompRend}). This causes the bridge arm to have a deficit in star forming gas compared to the tail arm, which would result in the asymmetric star formation properties seen in Figure \ref{SFRarms}.

\subsubsection{Shocks and offsets}
\label{Sec:offset}
We now look at the offset of gas and stellar structures in these simulations. To do so we fit logarithmic spiral functions to the Live and Pert simulations at multiple time frames, which is a function of the form:
\begin{equation}
\theta = \ln{(R/R_o)} {\rm cot}(\alpha)
\label{PitchEq}
\end{equation}
where the pitch angle is $\alpha$, and the parameter $R_o$ simply defines the initial azimuthal position of the spiral at $r=0$. We fit arms using the same method as described in Paper\,1, using a combination of Fourier decomposition and a fitting routine to constrain equation\,\ref{PitchEq} to the simulated arms. 

\begin{figure}
\centering
\resizebox{1.\hsize}{!}{\includegraphics[trim = 20mm 8mm 20mm 10mm]{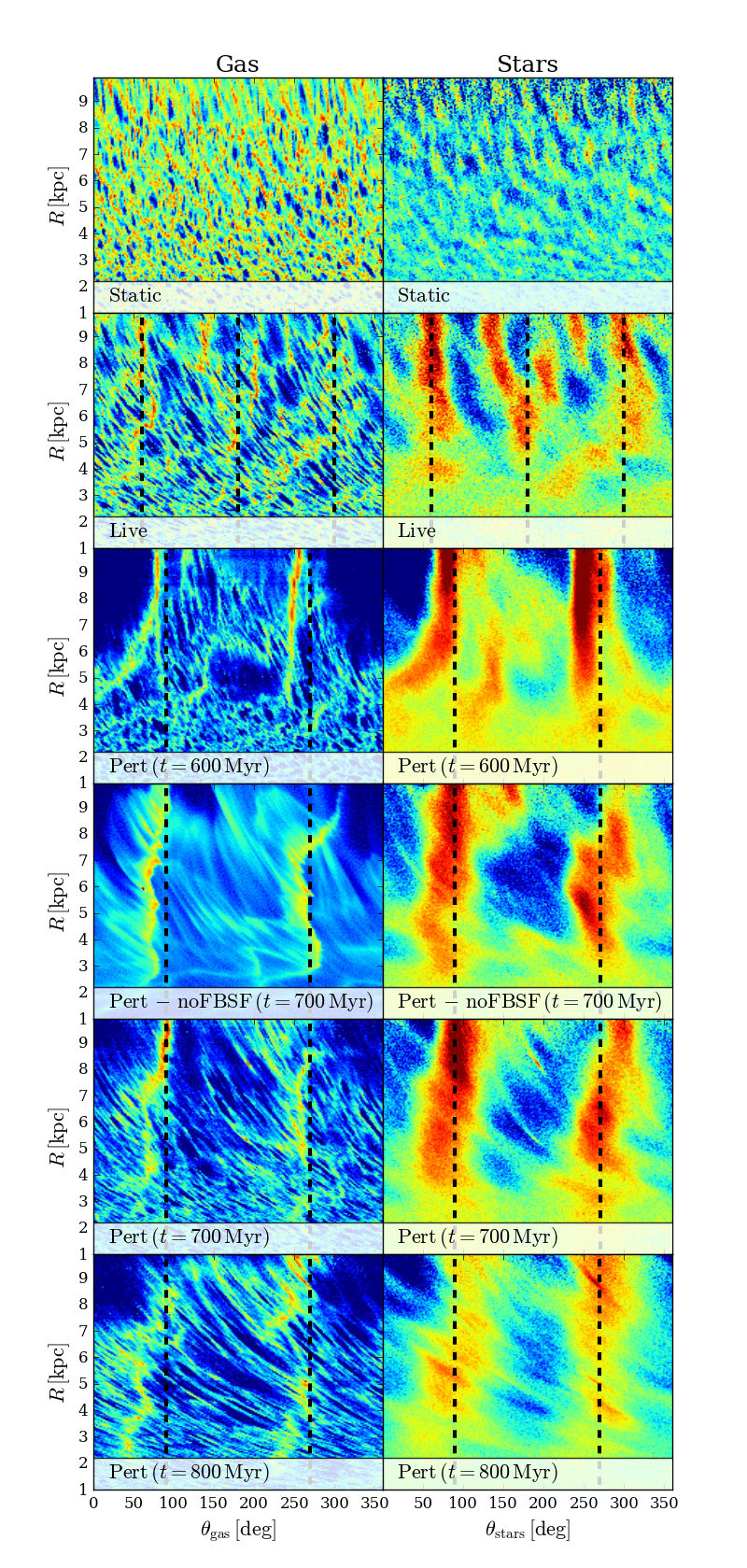}}
\caption{The spiral arms folded into $R-\theta$ space in the Static, Live, Pert and Pert-noFBSF calculations. On the left are the gaseous features, and on the right are the stars. Three time-stamps of the Pert calculation have been included, separated by 100Myr. For Live and Pert models, material has been aligned with a fit to the stellar spiral arms, with black dashed lines showing the fitted arm minima.}
\label{AzimuthBinning}
\end{figure}

In the Live simulation, with feedback and star formation at the same time as Figure \ref{LiveRenders}, we find the dominant mode to be that of $m=3$ and pitch angle of $\alpha=18^\circ$. The exact values tend to fluctuate somewhat with time, but a spiral feature is persistent approximately half a rotation until it is wound up and sheared away. The perturbed stellar disc behaves similarly to Paper\,1, in that a strong $m=2$ spiral starts off very wide and winds up of the order of a few rotations. For the time frame of the perturbed disc shown in the following figures, the pitch angle is approximately $\alpha=16^\circ$ at 700Myr and $\alpha=11^\circ$ at 800Myr.

\begin{figure}
\centering
\resizebox{0.6\hsize}{!}{\includegraphics[trim = 10mm 10mm 20mm 10mm]{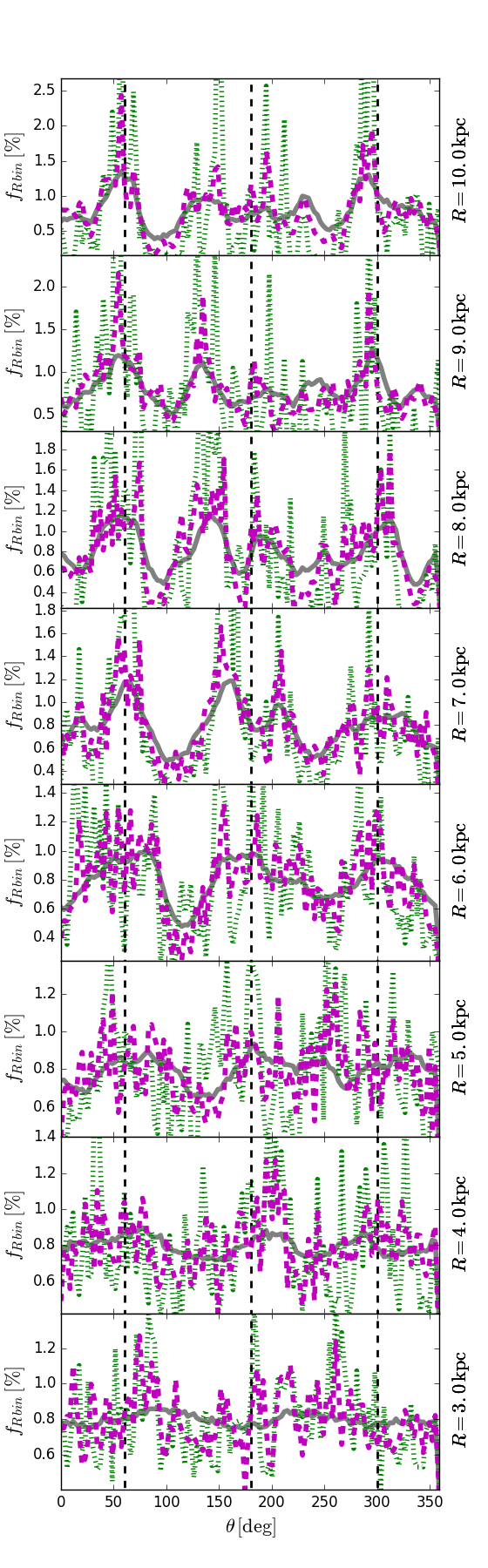}}
\caption{Location of old stellar (solid grey), gaseous (dashed magenta) and young stellar (dotted green) features in the Live galactic disc simulation and 600Myr of evolution. The $y$-axis shows the fraction of particles as a function of azimuth that are present in the denoted annulus; $f_{Rbin}$. Material has been phase-shifted to align to the fitted stellar spiral arms (black dashed lines).}
\label{OffISO}
\end{figure}

\begin{figure}
\centering
\resizebox{0.6\hsize}{!}{\includegraphics[trim = 10mm 10mm 20mm 10mm]{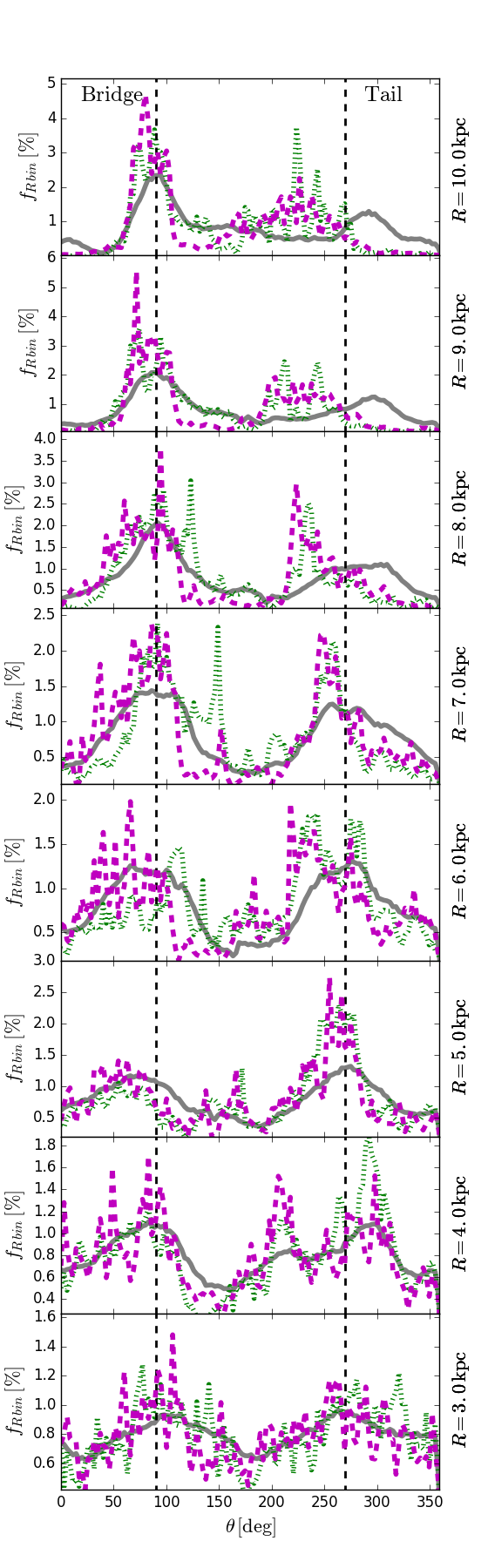}}
\caption{As Fig. \ref{OffISO} but for the perturbed stellar disc after 700Myr of evolution. Note that peaks are slightly offset between the different media compared to the arms in Fig. \ref{OffISO}. The right hand dashed line is the tail (the large offset at 10kpc), and the left hand dashed line is the bridge.}
\label{OffPERT}
\end{figure}

In Figure \ref{AzimuthBinning} we show the gaseous (left-hand column) and stellar (right-hand column, encompassing all ages) material from the Stat (top row), Live (second row) and Pert (third and below) simulations. We show three time-stamps of the perturbed simulation as the morphology is highly time-dependent. In the fourth row we show the 700Myr time-stamp for the Pert simulation without cooling, star formation, and feedback (noFBSF). For the Live and Pert models, material has been shifted in azimuth to align with the spiral arm fit of the old stellar population in each simulation, with the vertical black lines signifying the log-spiral arm fit. The Stat model has not been shifted due to a lack of prevailing spiral mode, showing a clear flocculent morphology in the these projections in both the stars and the gas. Bins have been then normalised to the azimuthally averaged surface density at each radius to remove the impact of the exponentially radial decay of the material and highlight overdensities. While clearly the old stars do not trace the log-spiral perfectly at all radii, the function provides a good match for the mid disc in all cases. The Live disc in particular is not entirely axisymmetric of the order of the fitted spiral, and there was also significant power in other modes at this time. 
The dominant arm number increases with radius in the stellar disc, with an $m=2$ mode around $R=3$kpc, $m=3$ at $R=5$kpc, $m=4$ at $R=7.5$kpc and an $m=5$ feature approaching $R=10$kpc\footnote{This is in accordance with predictions of swing amplification theory for axisymmetric perturbations, where the predicted spiral arm number is given by $m\approx \kappa^2R/4\pi G \Sigma$, where $\kappa$ is the epicycle frequency \citep{1981seng.proc..111T,1984ApJ...282...61S,2003MNRAS.344..358B}, see \citet{2014PASA...31...35D} Sections 2.1.3 and 2.2.1 for a more contemporary discussion.}
The Pert disc shows $m=2$ dominating at all times, at most radii in the disc, with the gas showing large cavities in the interarm regions compared to the Live case. The 600Myr time stamp of the Pert simulation shows residual $m=3$ mode signatures in the mid-disc, the same features as in the unperturbed Live calculation. Note that small spurs/feather like features can be seen in the Live and Pert gas discs. However, the ones in the Live run are significantly smaller features, and extrude at a shallower angle compared to the spiral arms. The spurs in the Pert disc appear larger and at a near-tangental angle to the spiral arms in some cases. The Pert and Pert-noFBSF runs have very similar morphology, with interarm branches appearing in the same position in both.

We plot the azimuthal position of the particles binned into annuli of 1kpc width in Figure \ref{OffISO} for the isolated live disc, and Figure \ref{OffPERT} for the perturbed disc at 700Myr. Each vertical panel shows different radii, and the black dashed lines are again the fitted spiral arms. The dashed magenta lines show the gas, while the grey and dotted green lines show the old and newly formed stars respectively (chosen here as $t_{\rm age} \leq 40{\rm Myr}$). Note that for the perturbed disc we have masked out two large clumps in the disc for clarity. These clumps can be clearly seen in Figure \ref{PertRenders}, giving clear narrow spikes in the binned distributions, and are dynamically decoupled from the spiral arms. We only plot data in the range $3{\rm kpc}<R<10{\rm kpc}$, as there are only weak spiral features further inward, and particle density drops considerably further out.

The Live stellar disc simulation shown in figures \ref{AzimuthBinning} and \ref{OffISO} has no strong offset between any of the media, with gas, stellar and young stellar peaks seemingly coinciding at all radii. This is also seen to be the case in other studies of live discs (e.g. \citealt{2011ApJ...735....1W,2012MNRAS.426..167G,2015PASJ...67L...4B}). The gas and young stars tend to show a greater arm-interarm contrast compared to the old stellar arms. In the inner disc (3-4kpc) there appears to be some minor offset between the stars and gas, though inspection of Figure \ref{AzimuthBinning} shows the arm features are extremely weak at this radius compared to the mid and outer disc.

\begin{figure}
\centering
\resizebox{1.\hsize}{!}{\includegraphics[trim = 10mm 10mm 10mm 0mm]{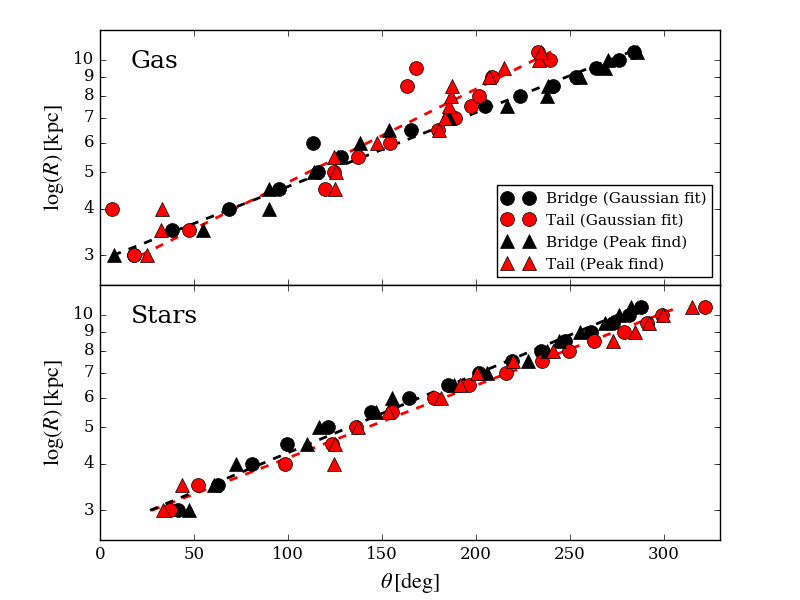}}
\caption{Tracing of the arms in the perturbed spiral simulation at 700Myr. The top panel shows the gaseous arms, and the bottom the stellar arms. Black points show the bridge arm, and red points the tail arm. Triangles denote arm finding by simply picking the peak, and circles by fitting a Gaussian to arm sections. Dashed lines show log-spiral fits to the bridge and tail arms for the peak-finding method (results were similar using the Gaussian fitting so for clarity a line is not shown).} 
\label{ArmTrace}
\end{figure}

The Pert simulation however appears to show some asymmetries between the different disc material. Even by-eye inspection of the particle binning in Figure \ref{AzimuthBinning} is enough to see that there are some positions where the gas and stars have a non-zero offset (e.g. $3{\rm kpc}<R<7{\rm kpc}$ in the 700Myr time frame). Inspection of Figure\,\ref{OffPERT} shows the offsets clearly, especially in the outer disc where the induced spiral is stronger. The offset appears to be strongest in the tail arm feature. The young stars tend to behave very similar to the gas, with only a young clump at 7kpc as the main outlier. As the definition of `young' stars is made to encompass a wider age range, the young stars tend to trace the same features as the old (i.e. present from $t=0$Myr) stellar population, whereas as the range is narrowed they coincide more with the gas arms.

\begin{figure*}
\centering
\resizebox{0.45\hsize}{!}{\includegraphics[trim = 01mm 0mm 00mm 0mm]{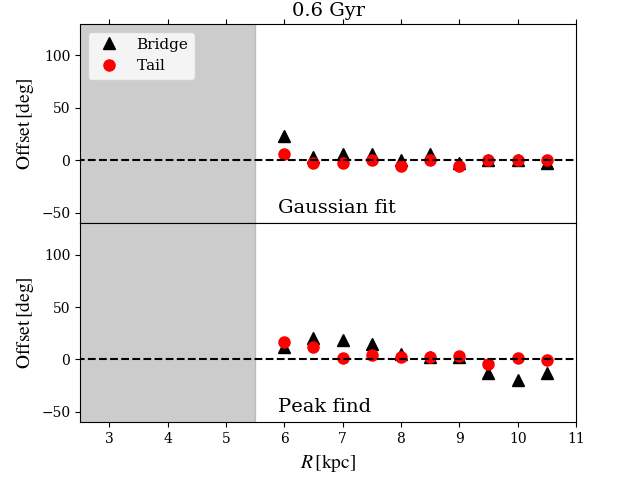}}
\resizebox{0.45\hsize}{!}{\includegraphics[trim = 00mm 0mm 00mm 0mm]{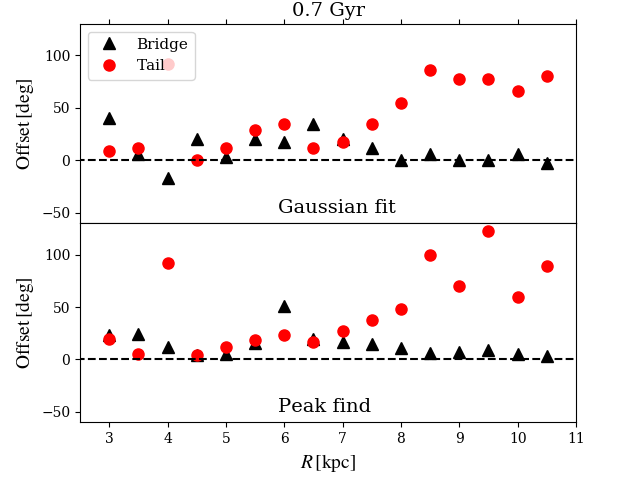}}
\resizebox{0.45\hsize}{!}{\includegraphics[trim = 01mm 0mm 00mm 0mm]{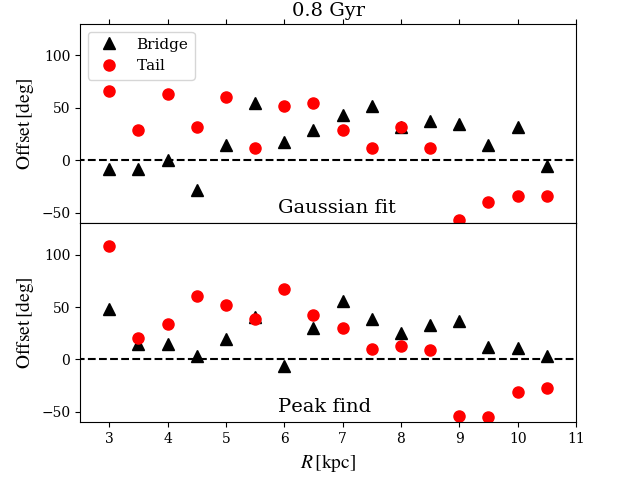}}
\resizebox{0.45\hsize}{!}{\includegraphics[trim = 00mm 0mm 00mm 0mm]{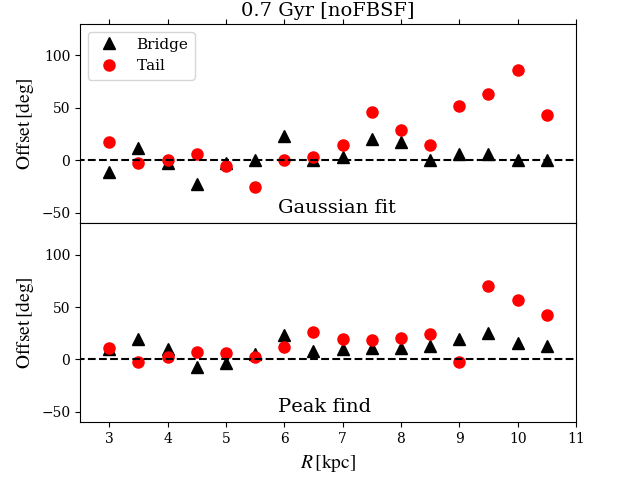}}
\caption{The offset between gaseous and stellar arms using the Gaussian fitting (top panels) and peak finding (bottom panels) methods in four different simulation time-stamps. Red circles show tail arm offsets, and black triangles the bridge arm offsets. The bottom right plot shows the same time-frame as the top right (0.7Gyr) but for a calculation without cooling, star formation and feedback. The shaded grey regions indicate radii where two-armed tidal spirals were not the dominant arm feature. Gas on the concave side of the stellar arms is defined as having negative offset (as upstream is defined in \citealt{2017MNRAS.465..460E}).}
\label{ArmOffset}
\end{figure*}

To further hone-in on this offset, we follow an analysis similar to that of \citet{2017MNRAS.465..460E}, who analyse the spiral arm offsets in the grand design galaxy M51. In their work the authors trace spirals in the gas and stars across the disc and measure their azimuthal positions relative to each other. We fit the spiral features using both of their approaches. The first is a simple peak-finding method in the vicinity of the spiral (i.e. in the angular ranges of $0^\circ\leq \theta < 180^\circ$ and $180^\circ \leq \theta < 360^\circ$ in Figure\,\ref{OffPERT}), while the second involves fitting a Gaussian function and taking the centre of the Gaussian as the arm location. The results of this analysis for a single time-frame (0.7Gyr in the Pert model) are shown in Figure\,\ref{ArmTrace}, where arms have been realigned to their true azimuth and binned into a 0.5kpc scale. The figure shows gas and stellar arm traces, as well as the two methods of arm location. All arm features clearly trace log-sprials (appearing as straight lines in $\log(R)-\theta$ space). The bridge and tail are slightly different at this epoch, with the tail being tighter wound than the bridge arm in the stars but more unwound than the bridge in the gas. The tail also shows much greater scatter than the bridge in the gas, which is likely due to the large spurring and strong radial decay of the arm at large radii (see the 0.7Gyr time-stamp in Fig. \ref{PertRenders}). The difference between the peak and Gaussian fitting techniques is only minor in each case, which was also concluded by \citet{2017MNRAS.465..460E}. The dashed lines show fits to the log-spiral function. Only the fits to the peak-finding points are shown, as there was only minor difference to the Gaussian fit. The pitch angles measured for the gas arms are $\alpha=15^\circ$ for the bridge and $\alpha=18^\circ$ for the tail, whereas the stars are best fitted by $\alpha=15^\circ$ for the bridge and $\alpha=14^\circ$ for the tail.

Offsets between stellar and gaseous arms are shown in Figure \ref{ArmOffset}, simply calculated from the difference between the points in Figure \ref{ArmTrace}. Here, we show three time-stamps of the Pert simulation (0.6, 0.7 and 0.8Gyr) and one of the noFBSF case (0.7 Gyr). The grey shaded region in the 0.6Gyr time-stamp shows where a $m=2$ spiral could not be fit, as the spiral had not yet reached the inner disc. The earliest time frame (0.6Gyr) shows that no strong offset between the stars and the gas is evident. There is a slight positive offset in the inner disc (where gas is on the convex side of the stellar spiral) and negative at outer radii (gas on the concave side), but the latter is only for the peak-find method. The positive offset component can be seen in Figure \ref{PertEvol} where the southern arm has strayed from a log-spiral in the gas pointing directly to the perturber. At 0.7Gyr there is a clear positive offset at almost all radii. In the inner and mid-disc this offset is in the range of 5\arcdeg{}- 25\arcdeg{}. The extremely large offsets in the outer disc are due to the gas tail branching at around 9kpc (seen in Figures \ref{PertRenders}, \ref{SFRarmstop} and \ref{AzimuthBinning} at 700Myr). The stellar arm continues to trace a log-spiral structure at this radii, whereas the gas deviates away to a much wider arm. At later times this wide branch dissipates, leaving only the shorter gas spiral that follows the stellar arm much closer. Only minor differences are seen between the different arm definitions. The noFBSF calculation shows similar features to the more sophisticated model, with a preference to positive offsets that is stronger at outer radii. In general the offsets appear slightly weaker in this hydrodynamic+gravity only run.

At the later time-stamp (0.8Gyr), the offsets show a much greater scatter with radius, though still showing a general preference to gas being on the convex side of the stellar arms. The binned data also highlight this (bottom panels Fig. \ref{AzimuthBinning}), with several arm components appearing offset in the gas compared to the stars. There is a minor negative offset component in the tail arm at a large radius. Inspection of top-down maps shows that in this region the tail arm is somewhat disconnected from the mid and inner disc (also visible in Fig. \ref{AzimuthBinning}) and is a combination of a few particularly large spurs/branches that are leaving the bridge arm and falling into the stellar potential well. This can be seen in the southern arm at 0.8Gyr in Figure \ref{PertEvol} where the arm seems to be winding inwards to a couple of large spurs breaking away from the northern bridge arm.

\begin{figure}
\centering
\resizebox{1.\hsize}{!}{\includegraphics[trim = 10mm 10mm 15mm 5mm]{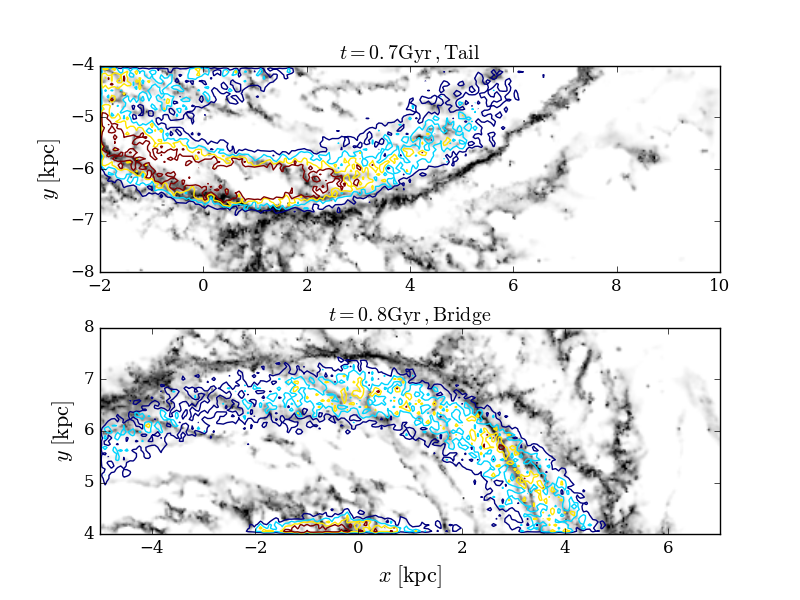}}
\caption{Two arm regions showing clear offsets in the gas (grey-scale render), and stars (coloured contours) at two different times (0.7 and 0.8 Gyr). The earlier time frame focuses on the offset in the tail, and the latter the bridge. Notice the weaker stellar surface density in the bridge arm at this time. Both images have been rotated in $z$-axis so that the arms lie horizontally.}
\label{Offsetall}
\end{figure}

\begin{figure}
\centering
\resizebox{1.\hsize}{!}{\includegraphics[trim = 0mm 10mm 0mm 10mm]{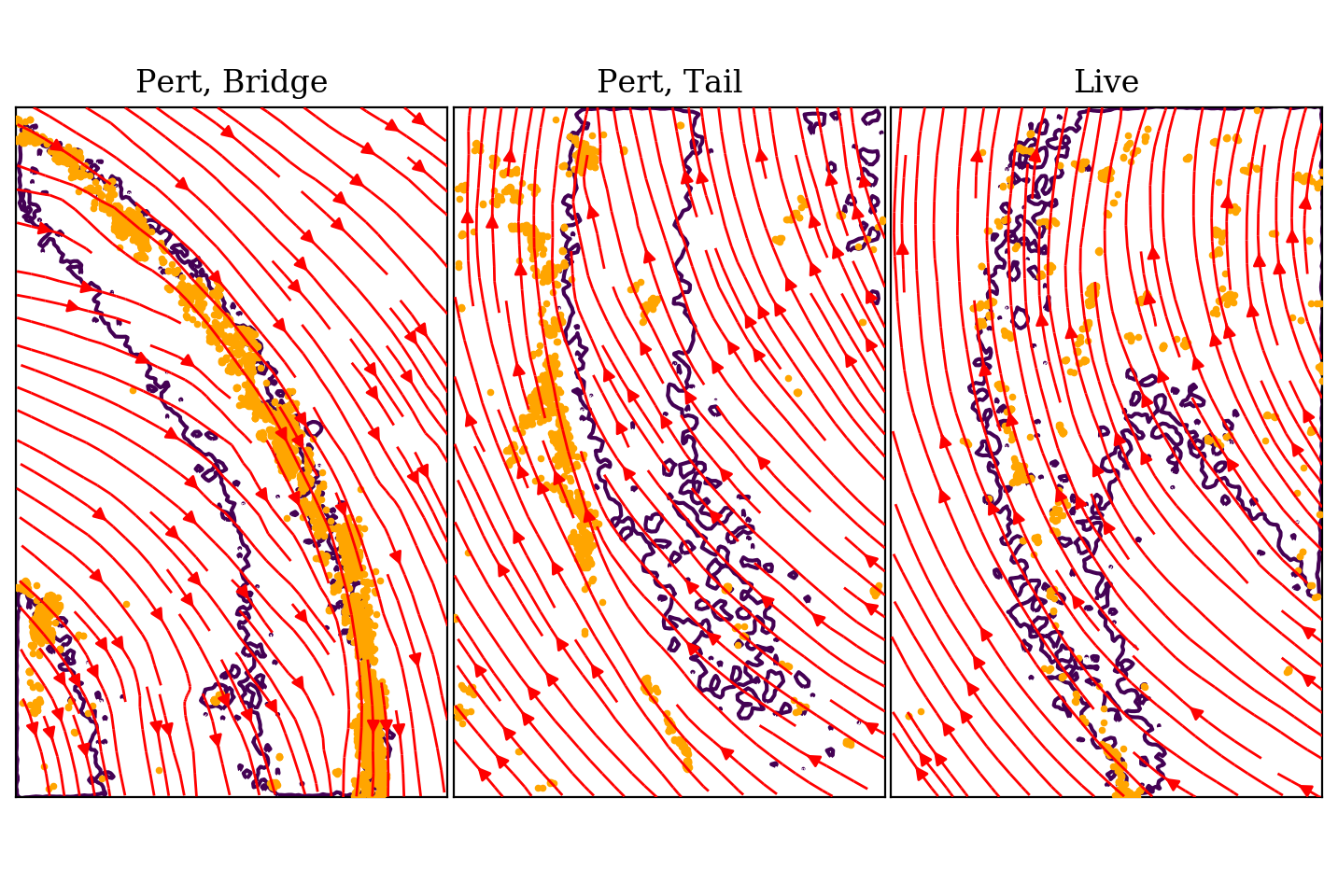}}
\caption{Plot of velocity fields around different spiral arms. From left to right: the bridge in a perturbed run, the tail in the same run, and an arm in the isolated live disc. Contours show the stellar density, and the orange points the regions of star formation. The stellar density field contour for the Live data has a value half that of the perturbed run due to the weaker nature of the spiral arms. Plots have dimensions of approximately 7kpc by 4kpc.}
\label{VelFields}
\end{figure}

In general, these models show the gas arms lying on the convex side of the stellar arms approximately 300Myr after the closest approach, with each arm showing a slightly different response, particularly in the outer disc region. Figure \ref{Offsetall} shows just how strong these offsets appear between the two different media, with stars (contours) and gas (render) showing a visible offset in two different arms at two different locations. In Figure \ref{VelFields}, we show a selection of stellar spiral arms (contours) with overlaying stellar velocity streamlines, as well as the star formation front (stars with ages $<20$Myr) as orange points. 
The first two panels show the same model as the 0.7Gyr model of Figure \ref{Offsetall} but now showing both bridge and tail arms, and the far right-hand panel shows an arm from the Live model (the same time frame as Figure \ref{LiveRenders}). It can clearly be seen that the velocity field in the perturbed spiral arms has regions of orbital crowding, with streamlines indicating the gas and star formation pile-up at the outside edge of the stellar spiral arms. Streamlines transition from an oblique angle to the arms, and then parallel to the arms after passing through. The star forming events are clustered along this pile-up zone, clearly offset from the stellar arms. The arms in the isolated stellar disc instead appear to leave the global velocity field relatively unperturbed, with no clear changes in direction after passing through the arms. The star forming regions in this arm seem uniformly scattered around the stellar arm, with no well defined formation front.

Comparing to other numerical studies; a similar convex-side offset is also seen by \citet{2008ApJ...675..188W}, and in the $<6$kpc region of the feedback/star formation model of \citet{2015PASJ...67L...4B}. However, both of these models used a density wave potential, which has clear corotation regions unlike these tidal arms and should result in different shock locations and offsets with radius. The best measurements of spiral arm offsets in observations come from M51. Recently, \citet{2017MNRAS.465..460E} looked into offsets between stellar and gaseous arms, seeing density-wave like offsets in the bridge arm (their `arm2') in the inner disc, showing a positive offset of the gas arm (downstream assuming their corotation placement). They also see negative offsets in a number of locations for the tail arm (their `arm1'), though there is a considerable scatter in their measured points. They also claim large uncertainties in outer-disc data, which is where our offsets are greatest. It is, however, somewhat problematic to compare our model with M51 directly, as the orbital configuration and mass models have not been tailored to the M51 system in any way. A more consistent comparison would come of using the initial conditions of \citet{2000MNRAS.319..377S,2010MNRAS.403..625D} or \citet{2011ApJ...743...35C} with the physics and resolution used here.

\subsubsection{Interarm spurs}

\begin{figure*}
\centering
\resizebox{0.77\hsize}{!}{\includegraphics[trim = 5mm 5mm 0mm 15mm]{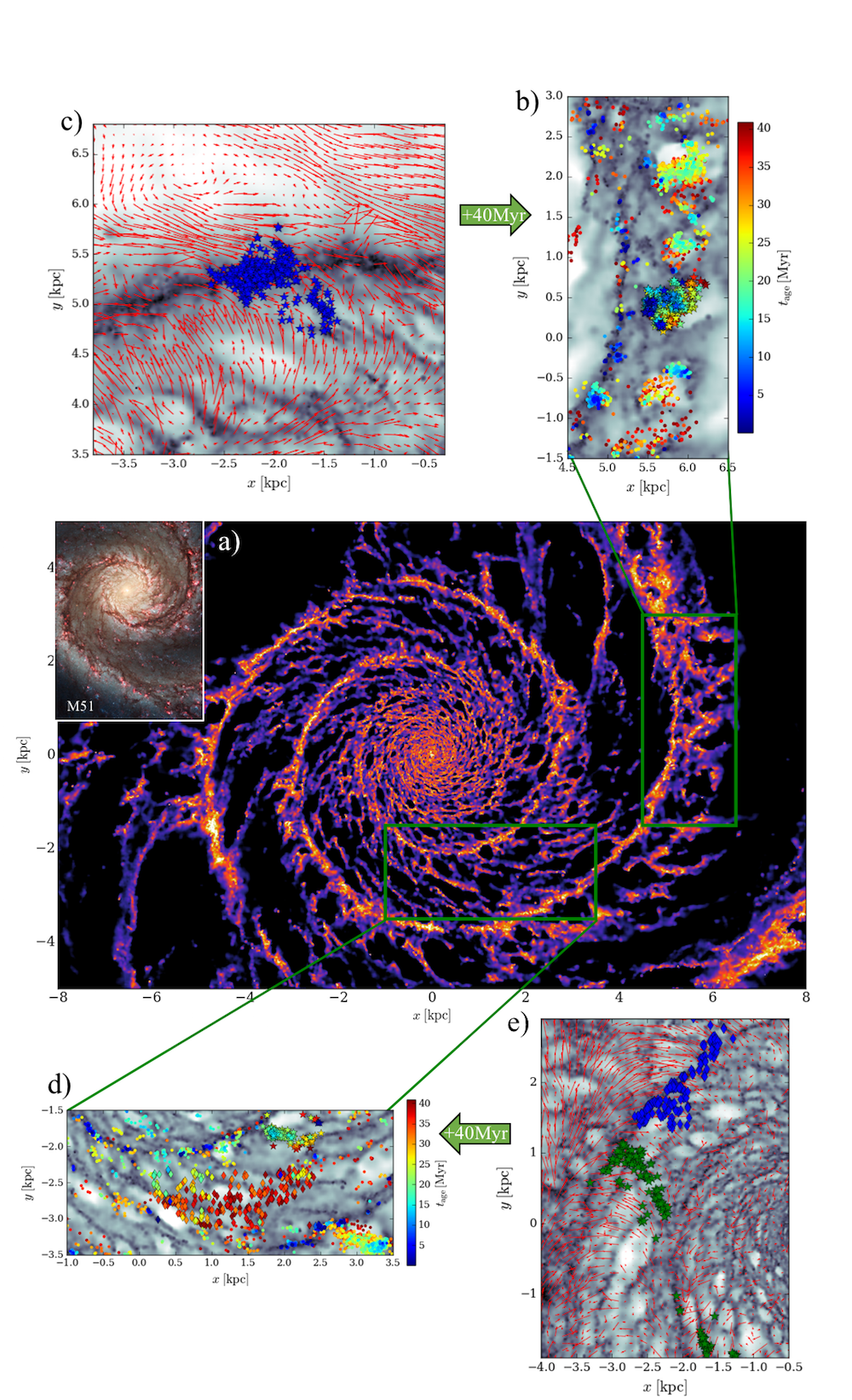}}
\caption{
A focus on spurs in the perturbed galactic disc.
Panel a) shows the large scale structure of the gas disc, with M51 shown in the insert for comparison of similar tidal spirals in nature.
Panels b) and d) show two regions of the disc with spur-like features, where young ($<$40Myr) star particles are overplotted as symbols coloured by their ages. Panels c) and e) show the starred and diamond regions of b) and d) 40Myr earlier, prior to the birth of these young stars. In c) and e) the corresponding parent gas particles to the new stars, shown as diamond and starred points in c) and d) are shown as coloured symbols. The local gas velocity field is shown by the red vector arrows.
M51 image credit: NASA, ESA, S. Beckwith (STScI) and The Hubble Heritage Team (STScI/AURA).}
\label{Spurs}
\end{figure*}

Spur features have been seen in numerous simulations of spiral galaxies, and yet their origin is a topic of debate. \citet{2004MNRAS.349..270W} propose that these features are a result of Kelvin-Helmholtz instabilities, whereas \citet{2006MNRAS.367..873D} interpret these features simply as the orbit crowding of gas that is then sheared out by differential rotation (see also \citealt{2006ApJ...646..213K}, \citealt{2006ApJ...647..997S}, \citealt{2014ApJ...789...68K}). Figure\;\ref{Spurs} shows an image of the gas in the perturbed disc with two different spur structures highlighted (panel a), one in the bridge and one in the tail. Clear spurring like features are seen, with those of the outer arm being stronger than the inner. Spirals of M51 are shown in the insert, where spurring features are also seen. The spurs here are uneven in nature, both within in an arm and between the bridge and tail. The spurs appear much larger in the bridge (left hand) arm, which is currently the arm with the dominant star formation activity (see Fig. \ref{SFRarms}). Some of the largest spurs even connect between the arms, for example the long branch-like features to the left of the eastern spur ($x\approx 4$kpc, $y\approx 3$kpc) that appears as a result of a large gas clump being between the tail arm it just left and the approaching bridge arm.

While the gas shows clear spurring features, it is worth mentioning at what form does the stellar distribution take. The old stellar population is very smooth, but the young stars show a level of substructure. Panels b) and d) show the spurred simulation regions again, now zoomed in and showing the young stellar population as coloured points. We defined young stars here as having ages $<$40Myr, and the points are coloured by their ages. In the b) arm, the youngest (bluer) stars are seen being born on the spiral arm itself, while the intermediate aged stars are flowing out of the spiral in clear clusters. The feedback from some of these clusters appears to be enforcing the cavities between spurs, as they appear coincident with voids in the gas. In the d) region the spurs are much weaker but more uniform in shape, showing segments a few kpc long. Despite some high density gas deposits in the spur regions, the youngest stars lie almost entirely along the spiral arm trough, with the older stars occupying most of the interarm region. These types of structures are sometimes denoted ``feathers" in the literature as they have little or no star formation, whereas spurs are defined as short, high density regions with active star formation (e.g. \citealt{2013MNRAS.436.1836R}).

In panels c) and e) we show the parent gas particles of the young stars seen in b) and d), showing the structure of the gas that birthed the young stars found in the spur regions. These data is shown 40Myr in the past of b) and d), far enough in the past as to be prior to the birth of the oldest ``young" star. We have only plotted a subset of the stars in b) and d), chosen to highlight what appear to be unique features. In b), these are the points indicated by star symbols, which are born from the blue starred gas particles in c). In d), two regions were selected. One is the older interarm region (diamond symbols) with stars born from the blue diamonds in e). The other is a short intermediate aged spur (star symbols in the top right of d) with stars born from the green starred gas particles in e). A grey-scale of the gas density is shown in the background, with additional red vector arrows indicating the rotation-subtracted velocity field of the gas. In both c) and e) the velocity field shows material entering the arms from both directions, though the gas on the outside edge of the arm is moving nearly parallel to the arm, whereas the gas on the inside is flowing in almost perpendicular to the arm. Over the 40Myr period all three subsets of young stars seem to show a similar structure in their parent gas particles as they do after passing through the spiral. For instance, the blue diamonds in e) show an elongated branch while the green starred points appear as a short upstream spur in gas that then transforms into a short downstream spur. 

Figure \ref{PertEvol} shows that spur features are evident at most times after the interaction, though become increasingly difficult to identify after the 1Gyr time-stamp due to the complex morphology and cavities bored out by the two high-density clumps. Inspection of the isolated spiral in Fig. \ref{LiveRenders} shows no clear spur features on the scale of the perturbed disc. There are some very small interarm features, but they appear to be the result of clustered supernova cavities being sheared out by rotation. This is similar to the Static disc case (Fig. \ref{StatRenders}) though as there is no clear stellar structure, it is difficult to define quite what is meant by a spur in this model.

\section{Conclusions}
\label{conclusions}

In this work we have presented a study of the gas morphology and star formation process in a tidally induced spiral galaxy using simulations including cooling, star formation and stellar feedback. The tidal encounter greatly boosts the star formation activity in the disc, and causes bursts of star formation in the bridge and tail spiral arms. The two arms exhibit a temporal offset in their bursts of star formation. Activity is first being boosted in the tail arm then the bridge arm, the result of gas stripping by the companion passage delaying the star formation in the bridge and extreme orbit overlapping in the tail arm causing rapid gas compression. 

The gaseous arms appear offset azimuthally compared to stellar arms, an effect not seen in the dynamic arms seen pre-interaction. These offsets are different between each arm, and appear to be highly time dependent. They do however show a clear preference for gas arms to be on the convex side of the stellar spiral potential well. Finally, we also observe spurring features in the interarm regions, similar to those seen in density wave spiral simulations and observations. Inner spurs appear the result of purely shearing motion in high density gas in the arms, whereas the presence of outer spurs is exemplified by clustered feedback by newly formed stars leaving the spiral arm.

These findings further separate tidal spirals from the other spiral generation mechanisms in the literature; providing valuable diagnostics for observations of the ISM in external galaxies. With the resolution of current surveys, such diagnostics are out of reach for all but the nearest spiral galaxies. Yet they will become increasingly important as we move closer towards pc scale observations and in turn a unified theory of spiral structure.

\section*{Acknowledgments}
We thank the referee, Curtis Struck, for his reading of this manuscript and insightful comments/suggestions that have improved this work. Numerical computations were (in part) carried out on Cray XC30 at Center for Computational Astrophysics, National Astronomical Observatory of Japan and the GPC supercomputer at the SciNet HPC Consortium \citep{2010JPhCS.256a2026L}. SciNet is funded by: the Canada Foundation for Innovation under the auspices of Compute Canada; the Government of Ontario; Ontario Research Fund - Research Excellence; and the University of Toronto. SMB acknowledges support from the Vanier Canada Graduate Scholarship Program. We used the \textsc{pynbody} \citep{2013ascl.soft05002P} and \textsc{yt} \citep{2011ApJS..192....9T} packages for parts of our analysis and plotting of data. We thank T. Okamoto and T. Saitoh for helpful discussions on this work.

\bibliographystyle{mnras}
\bibliography{Pettitt_armphys.bbl}


\bsp
\label{lastpage}
\end{document}